\documentclass[fleqn,usenatbib]{mnras}

\usepackage{newtxtext,newtxmath}

\usepackage[T1]{fontenc}

\DeclareRobustCommand{\VAN}[3]{#2}
\let\VANthebibliography\thebibliography
\def\thebibliography{\DeclareRobustCommand{\VAN}[3]{##3}\VANthebibliography}

\usepackage{graphicx}	
\usepackage{amsmath}	
\usepackage{booktabs}
\usepackage{amsfonts}
\usepackage{amscd}
\usepackage{physics}
\usepackage{natbib}
\usepackage{tikz}
\usetikzlibrary{shapes,arrows}
\usepackage[flushleft]{threeparttable}
\usepackage{orcidlink}

\newcommand{\lmax}{\lambda_{\mathrm{max}}}
\newcommand{\lmin}{\lambda_{\mathrm{min}}}
\newcommand{\gpcm}{\, \mathrm{g} \, \mathrm{cm}^{-3}}
\newcommand{\ergpcm}{\, \mathrm{erg} \, \mathrm{cm}^{-3}}
\newcommand{\ar}{a_{\mathrm{R}}}

\defcitealias{Skinner_2013}{SO13}
\defcitealias{Jiang_2013}{J13}

\title[VETTAM RHD Code]{VETTAM: A scheme for radiation hydrodynamics with adaptive mesh refinement using the variable Eddington tensor method}

\author[S.~H.~Menon et al.]{
Shyam H.~Menon$^{\orcidlink{0000-0001-5944-291X}}$,$^{1}$\thanks{E-mail: shyam.menon@anu.edu.au (SHM)}
Christoph Federrath$^{\orcidlink{0000-0002-0706-2306}}$,$^{1,2}$\thanks{E-mail: christoph.federrath@anu.edu.au (CF)}
Mark R.~Krumholz$^{\orcidlink{0000-0003-3893-854X}}$,$^{1,2}$
Rolf Kuiper$^{\orcidlink{0000-0003-2309-8963}}$,$^{3}$ \newauthor Benjamin D. Wibking$^{\orcidlink{0000-0003-3175-2291}}$,$^{1,2}$ and Manuel Jung$^{\orcidlink{0000-0001-8559-7544}}$ $^{4}$
\\
$^{1}$Research School of Astronomy and Astrophysics, Australian National University, Canberra, ACT~2611, Australia\\
$^{2}$ARC Centre of Excellence for Astronomy in Three Dimensions (ASTRO-3D), Canberra, ACT~2611, Australia\\
$^{3}$Zentrum f{\"u}r Astronomie der Universit{\"a}t Heidelberg, Institut f{\"u}r Theoretische Astrophysik, Albert-Ueberle-Stra{\ss}e 2, 69120 Heidelberg, Germany\\
$^{4}$Hamburger Sternwarte, Universit{\"a}t Hamburg, Gojenbergsweg 112, 21029 Hamburg, Germany
}

\date{Accepted XXX. Received YYY; in original form ZZZ}

\pubyear{2022}

\begin{document}
\label{firstpage}
\pagerange{\pageref{firstpage}--\pageref{lastpage}}
\maketitle

\begin{abstract}
We present Variable Eddington Tensor-closed Transport on Adaptive Meshes (\texttt{VETTAM}), a new algorithm to solve the equations of radiation hydrodynamics (RHD) with support for adaptive mesh refinement (AMR) in a frequency-integrated, two-moment formulation. The method is based on a non-local Variable Eddington Tensor (VET) closure computed
with a hybrid characteristics scheme for ray tracing. We use a Godunov method for the hyperbolic transport of radiation with an implicit backwards-Euler temporal update to avoid the explicit timestep constraint imposed by the light-crossing time, and a fixed-point Picard iteration scheme to handle the nonlinear gas-radiation exchange term, with the two implicit update stages jointly iterated to convergence. We also develop a modified wave-speed correction method for AMR, which we find to be crucial for obtaining accurate results in the diffusion regime. We demonstrate the robustness of our scheme with a suite of pure radiation and RHD tests, and show that it successfully captures
the streaming, static diffusion, and dynamic diffusion regimes and the spatial transitions between them, casts sharp shadows, and yields accurate results for rates of momentum and energy exchange between radiation and gas. A comparison between different closures for the radiation moment equations, with the Eddington approximation (0th-moment closure) and the $M_1$ approximation (1st-moment closure), demonstrates the advantages of the VET method (2nd-moment closure) over the simpler closure schemes.
\texttt{VETTAM} has been coupled to the AMR \texttt{FLASH} (magneto-)hydrodynamics code and we summarize by reporting performance features and bottlenecks of our implementation.
\end{abstract}




\section{Introduction}
Radiation hydrodynamics (RHD) plays a crucial role in the evolution of several astrophysical systems, such as stellar atmospheres \citep[e.g.,][]{Mihalas_1978,Castor_2004}, planetary atmospheres \citep[e.g.,][]{Zhang_2020}, stellar winds \citep[e.g.,][]{Castor_1975,Smith_2014}, circumstellar disks \citep[e.g.,][]{Turner_2014,Zhao_2020}, supernovae \citep[e.g.,][]{Janka_2007}, star-forming clouds in the interstellar medium \citep[e.g.,][]{Krumholz_2019,Rosen_2020}, active galactic nuclei and their associated jets \citep[e.g.,][]{Davis_2020}, and in galactic outflows \citep[e.g.,][]{Naab_2017,Zhang_2018}. These systems span a vast range of scales and physical conditions, which can be parameterised by the optical depth across the region of interest, determining how radiation is transported. While there exist various numerical techniques to solve the RHD equations in some limiting cases (e.g., high vs.~low optical depth), a crucial requirement of flexible modern methods is their ability to treat a mixture of regimes in a robust and efficient way.

There are well-known difficulties associated with numerically solving the RHD equations. The primary challenge has to do with the multidimensional nature of the radiation intensity -- a function of spatial location, time, direction and frequency -- that effectively makes the radiative transfer (RT) equation very expensive to solve, especially in a dynamical system where this needs to be done multiple times \citep[however, see][]{Jiang_2021}. A common approach to circumvent this is to integrate the RT equation over all frequencies and angles to obtain the gray radiation moment equations, reducing the dimensionality of the system \citep[e.g.,][]{Pomraning_1973,Mihalas_1984,Castor_2004}. However, this introduces the need for an extra closure equation to estimate the moments of the radiation intensity whose evolution is not explicitly computed. One commonly-used closure is the flux-limited diffusion (FLD) method \citep[e.g.,][]{Turner_2001,Howell_2003,Krumholz_2007a,Gittings_2008,Swesty_2009,Kuiper_2010,Commercon_2011,Zhang_2011,VanderHolst_2011, Flock_2013,Bitsch_2013,Klassen_2014,Ramsey_2015,Chatzopoulos_2019,Moens_2021}, which closes the equations at the first moment (the radiation flux), which is assumed to be proportional to the negative of the gradient in radiation energy density; this then implies that the Eddington tensor is locally isotropic \citep{Levermore_1981}. The FLD closure reduces the radiation transport to a parabolic diffusion equation, with a diffusion coefficient chosen to limit the photon speed to be smaller than the speed of light. However, this method often suffers from inaccuracies in the optically thin regime, or when a mixture of low- and high-opacity gas is present. For instance, FLD methods cannot cast shadows \citep{Hayes_2003, Kuiper_2013}.

A more accurate closure scheme that has recently been adopted widely is the $M_1$ closure \citep[e.g.,][]{Gonzalez_2007,Aubert_2008,Skinner_2013,Rosdahl_2013,Rosdahl_2015,Kannan_2019,Skinner_2019,Bloch_2021,Fuksman_2021,Chan_2021,Wibking_2021}, which retains the time evolution of the radiation flux and adopts a local closure relation for the radiation pressure tensor, or equivalently the Eddington tensor, in terms of the local radiation energy density and flux; a variety of assumptions regarding the nature of the radiation field are possible, each yielding slightly different versions of the closure relation \citep{Minerbo_1978,Levermore_1984}. While the $M_1$ closure can handle transitions in optical depths for a single beam of radiation, it fails for other non-trivial geometrical distributions of radiation sources. For instance, the presence of multiple sources interacting in optically thin media causes un-physical discontinuities between the sources' radiation fronts, and produces spurious fluxes in the direction perpendicular to the line connecting the sources.

A more accurate alternative is the so-called Variable Eddington Tensor (VET) scheme \citep[e.g.,][]{Stone_1992,Gehmeyr_1994,Sekora_2010,Jiang_2012,Asahina_2020}, a non-local scheme that does not adopt a closure relation or model \textit{a priori}, but rather computes the Eddington tensor self-consistently through a formal solution of the time-independent RT equation along discrete rays using a ray-tracing approach \citep[e.g.,][]{Davis_2012}. The self-consistently computed closure is combined with the radiation moment equations to solve for the radiation quantities. While more computationally expensive due to the required non-local ray-trace solution and its associated communication overheads, the VET approach does not face the shortcomings of the more approximate closure models discussed above. For example, the FLD and $M_1$ closure schemes can produce misleading results in some semitransparent problems \citep{Krumholz_2012,Kuiper_2013,Rosdahl_2015,Kannan_2019}, and this was discovered only through a comparison of these simpler methods with a VET scheme \citep{Davis_2014} or other non-local closure schemes such as a Monte Carlo approach \citep{Tsang_2015,Harries_2015,Smith_2020}. 

Another difficulty associated with RHD is the vast difference in scale between the wave speeds associated with radiation and hydrodynamics -- the speed of light ($c$) and gas flow speed ($v$)\footnote{For strongly subsonic flows, the relavant wavespeed for the timestep is the sound speed $c_\mathrm{s}$.} in the medium, respectively. In many non-relativistic, astrophysical problems, $v/c \ll 1$, but stability constraints associated with explicit temporal updates restrict the timestep to the stringent radiation transport timescale, which renders simulations that must follow the system for several hydrodynamical timesteps computationally infeasible. A popular approach to alleviate this issue is to adopt a reduced speed-of-light approximation (RSLA), wherein the speed of light is reduced to a fraction of its true value $\hat{c}$ \citep[e.g.,][]{Gnedin_2001, Skinner_2013}. This allows one to use an explicit temporal update for the radiation quantities that is now limited by the much larger ratio $c_\mathrm{s}/\hat{c}$. Explicit updates of the radiation moment equations with the RSLA have the advantage that they can adopt widely-studied and well-tested tools for solving hyperbolic PDEs. These schemes are also, in general, well parallelizable and scalable, and can be accelerated with GPUs \citep{Wibking_2021}. However, the RSLA is only applicable under the condition that the hierarchy of evolution timescales -- namely the light crossing, radiation diffusion, and hydrodynamic timescales -- remains the same even with the reduced speed of light. This places constraints on the applicability of the RSLA in systems with high optical depth, the dynamic diffusion regime, limiting the overall flexibility of the scheme.

An alternative approach is to treat the transport of radiation at the hydrodynamic timestep in a fully-implicit fashion to avoid stability-related issues. This approach has been used in a vast variety of implementations, and has the benefit of being applicable in a broad range of systems. However, implicit methods require the solution of a large system of equations using sparse matrix solvers, whose performance and scalability are limited. This is aggravated by the presence of stiff, nonlinear terms that need to be handled implicitly along with the radiation quantities, rendering the system of equations both non-local and nonlinear. However, implicit, non-local methods remain the gold standard of accuracy, and recent advancements in numerical methods, and the development of freely available and continually improving libraries of linear/nonlinear sparse matrix solvers using Krylov subspace methods, has accelerated the development of implicit RHD schemes that can be applied on massively parallel computing architectures \citep{Saad_2003}. 

In this paper, we present \texttt{VETTAM}\footnote{\texttt{VETTAM} is an acronym for Variable Eddington Tensor closed Transport on Adaptive Meshes. The acronym stands for \textit{light} in the south Indian language of Malayalam -- the first language of SHM.}, the first multidimensional radiation moment scheme closed using a VET computed through a formal solution of the RT equation with Adaptive Mesh Refinement (AMR) capabilities. The formal solution is computed through a method based on the hybrid characteristics ray-tracing implemented in \citet{Buntemeyer_2016}\footnote{The module has been rewritten and improved significantly by Manuel Jung.}. We couple this with the update of the radiation moment equations in a fully time-implicit fashion that can handle all regimes of optical depth in radiation transport. We solve the resulting implicit system of nonlinear equations with a fixed-point Picard iteration scheme that allows us to use a variety of sparse Krylov subspace matrix solvers provided with the \texttt{PETSc} library \footnote{\url{https://petsc.org/release/}} \citep{PetscConf,PetscRef}. We describe our scheme, its salient features, and its integration into the \texttt{FLASH} code \citep{Fryxell_2000, Dubey_2008} in Section~\ref{sec:Methods}. In Section~\ref{sec:Tests}, we present a comprehensive test suite to demonstrate the accuracy and capabilities of our algorithm. In Section~\ref{sec:Discussion} we discuss the advantages of our VET scheme over methods that adopt simpler closures, touch upon the performance capabilities of our scheme, and list some caveats associated with our implementation that provide scope for future improvements. We briefly summarise in Section~\ref{sec:Summary} and mention potential applications for which we intend to use \texttt{VETTAM} in the near future.

\section{Numerical Methodology}
\label{sec:Methods}
In this section, we describe our implementation to treat the coupled radiation hydrodynamic set of equations, fully integrated into the \texttt{FLASH} code \citep{Fryxell_2000, Dubey_2008}. \texttt{FLASH} is a publicly available high-performance general application (astro-)physics code that includes a wide range of physical capabilities and is designed in an organised modular fashion \citep{Dubey_2019}. It solves the hydrodynamic equations on an Eulerian mesh, with Adaptive Mesh Refinement \citep{Berger_1989} using the \texttt{PARAMESH} library \citep{Macneice_2000}. By default it uses a modified second derivative normalised by the average gradient of a chosen variable over a cell as a dimensionless criterion for refinement \citep{Lohner_1987}, although other refinement criteria are available or straightforward to implement \citep[such as Jeans refinement][]{Federrath_2010_Sinks}.

\subsection{Equations of Radiation Hydrodynamics}

In \texttt{VETTAM}, we solve the equations of non-relativistic gray (frequency-integrated) RHD in conservative form, written in the mixed-frame formulation, i.e., where the moments of the radiation intensity are written in the lab frame, and the opacities are written in the comoving frame, with the transformation between the frames accounted by $\mathcal{O}(v/c)$ terms in the resulting equations \citep[e.g.,][]{Mihalas_1982,Krumholz_2007a}. This approach takes advantage of the simplicity of the hyperbolic operators in the lab frame, and the simplicity of the matter emissivities and opacities in the comoving frame, where they are generally isotropic \citep[see][for a detailed critique of these various approaches]{Castor_2009}. An additional advantage is that the mixed-frame formulation permits conservation of total energy, whereas a comoving-frame formulation of the equations does not; this is especially important for AMR, where non-conservation may be amplified by repeated refinements. However, the $\mathcal{O}(v/c)$ expansions to transform the opacities in the mixed-frame equations fail for emission/absorption lines as soon as $v/c$ becomes comparable to $\Delta \lambda/\lambda_0$, where $\Delta\lambda$ and $\lambda_0$ are the linewidth and line centre; this limits the use of the mixed-frame approach to broad lines or continuum radiation. Since we are interested in modelling dust continuum radiation with \texttt{VETTAM} on an AMR grid, the advantages of the mixed-frame formulation outweigh its disadvantages. In writing out the equations, we are careful to retain terms that are of leading order in all regimes of RHD, using the scalings for various terms given in Table~1 of \citet{Krumholz_2007a}, to ensure that our scheme recovers the correct asymptotic limits \citep{Lowrie_1999}. We neglect scattering for simplicity; however, an extension to include scattering would be straightforward. Finally, we assume the matter is always in local thermodynamic equilibrium (LTE), though not necessarily in equilibrium with the radiation field, and we treat the material property coefficients as isotropic in the comoving frame.

We adopt the following convention to represent the RHD operations: tensor contractions over a single index with dots (e.g., $\mathbfit{a} \cdot \mathbfit{b}$), tensor contractions over two indices by colons (e.g., \mathbfss{A}:\mathbfss{B}), and tensor products of vectors without an operator symbol (e.g., $\mathbfit{a}\mathbfit{b}$). The equations solved by \texttt{VETTAM} are then 
\begin{gather}
    \label{eq:continuityeq}
	\frac{\partial \rho}{\partial t} + \nabla \cdot (\rho \textbf{v}) = 0 \\
	\frac{\partial (\rho\textbf{v})}{\partial t} + \nabla \cdot (\rho \textbf{v}\textbf{v}) = - \nabla P - \rho \nabla \Phi + \mathbfit{G} + \dot{{\mathbfit{p}_*}}  \\
	\label{eq:gasMom}
	\frac{\partial E}{\partial t} + \nabla \cdot [(E + P)\textbf{v}] = -\rho \textbf{v} \cdot \nabla \Phi + cG^0 + \dot{E_*} \\
	\label{eq:Erad}
	\frac{\partial E_r}{\partial t} + \nabla \cdot \mathbfit{F}_r = -cG^0\\
	\frac{\partial \mathbfit{F}_r}{\partial t} + \nabla \cdot (c^2E_r\mathbfss{T}) = -c^2\mathbfit{G},
	\label{eq:Frad}
	\end{gather}
where the pressure is given by the ideal gas law,
\begin{equation}
	P = \frac{\rho k_\mathrm{B} T}{\mu} , 
	\label{eq:gasPressure}
\end{equation}
and
\begin{equation}
    \begin{aligned}
    G^0 =& \rho \kappa_E E_r - \rho \kappa_P a_RT^4 + \rho \left(\kappa_F - 2\kappa_E \right) \frac{\mathbfit{v} \cdot \mathbfit{F}_r}{c^2} \\
    &+ \rho \left( \kappa_E - \kappa_F \right) \left[\frac{v^2}{c^2}E_r + \frac{\mathbfit{v}\mathbfit{v}}{c^2} :\mathbfss{P}_r  \right] ,
\end{aligned}
\end{equation}
and
\begin{equation}
\label{eq:G}
\mathbfit{G} = \rho \kappa_R \frac{\mathbfit{F}_r}{c} - \rho \kappa_R E_r\frac{\mathbfit{v}}{c} \cdot (\mathbfss{I} + \mathbfss{T}),
\end{equation}
are the time-like and space-like parts of the specific radiation four-force density for a direction-independent flux spectrum \citep{Mihalas_2001} to leading order in all regimes. 
In the above equations $\rho$ is the mass density, $P$ the gas thermal pressure, $\mathbfit{v}$ the gas velocity, $\Phi$ the gravitational potential, $T$ the gas temperature, $\mathbfss{I}$ the identity matrix, and $c$ the speed of light in vacuum. $E$ is the total gas energy density, given by
\begin{equation}
E = E_g + \frac{1}{2}\rho v^2 ,
\end{equation}
where $E_g$ is the gas internal energy density. In the ideal gas law, Equation~\ref{eq:gasPressure}, $k_\mathrm{B}$ is the Boltzmann constant and $\mu$ the mean particle mass. As for the radiation quantities, $E_r$ is the lab-frame radiation energy density, $\mathbfit{F}_r$ the lab-frame radiation momentum density, $\mathbfss{P}_r$ is the lab-frame radiation pressure tensor, and $a_R$ the radiation constant. The radiation moment quantities are related to the radiation intensity $I_r(\hat{\mathbfit{n}}_k,\nu)$ travelling in direction $\hat{\mathbfit{n}}_k$ by the relations 
\begin{gather}
\label{eq:angularmoment0}
c E_r=\int_{0}^{\infty} d \nu \int d \Omega\, I_r(\hat{\mathbfit{n}}_k, \nu) \\
\mathbfit{F}_r=\int_{0}^{\infty} d \nu \int d \Omega\, \hat{\mathbfit{n}}_k\,I_r(\hat{\mathbfit{n}}_k, \nu) \\
c \mathbfss{P}_r=\int_{0}^{\infty} d \nu \int d \Omega\, \hat{\mathbfit{n}}_k \hat{\mathbfit{n}}_k\,I_r(\hat{\mathbfit{n}}_k, \nu),
\label{eq:angularmoment2}
\end{gather}
where $d\Omega$ and $d\nu$ are the infinitesimal solid angle and lab-frame frequencies, respectively. The radiation closure relation is used to close the above system of equations, and is of the form
\begin{equation}
\label{eq:closure}
\mathbfss{P}_r = \mathbfss{T}E_r,
\end{equation}
where $\mathbfss{T}$ is the Eddington Tensor. We use an Eddington tensor directly calculated from angular quadratures of the frequency-averaged specific intensity $I_r(\hat{\mathbfit{n}}_k)$, using relations~(\ref{eq:angularmoment0}) and~(\ref{eq:angularmoment2}), where $I_r$ as a function of the spatial path length $s$ is calculated from a formal solution of the time-independent radiative transfer equation
\begin{equation}
    \label{eq:RTeq}
    \frac{\partial I_r}{\partial s} = \rho \kappa(S - I_r),
\end{equation}
where $S$ is the source function, which, for the purposes of modelling the emission from dust grains, we set equal to the frequency-integrated Planck function $B(T) = ca_RT^4/(4\pi)$. The expression in Equation~\ref{eq:RTeq} neglects scattering, and assumes that the dust emits and absorbs radiation in the comoving frame with the same  gray opacity $\kappa = \int_{0}^{\infty} \kappa\left(v_{0}\right) d\nu_0$, where $\kappa\left(v_{0}\right)$ is the material opacity at frequency $\nu_0$. In addition, we also ignore $\mathcal{O}(v/c)$ terms in this equation, which arise from the mixed-frame formulation, since we expect the contribution of these terms to the Eddington tensor to be relatively low. $\dot{E_*}$ and $\dot{\mathbfit{p}_*}$ denote the energy and momentum deposition rates to the gas by the direct radiation from point sources or sink particles \citep{Federrath_2010_Sinks}. We split this direct contribution from the diffuse radiation modelled by the VET, and treat it directly, using only a ray-tracer on rays originating at the sources\footnote{This is sometimes referred to as a hybrid radiation transfer, and we follow this terminology in this paper.} \citep[][]{Wolfire_1986,Murray_1994,Kuiper_2010,Kolb_2013,Flock_2013,Bitsch_2013,Klassen_2014,Ramsey_2015,Rosen_2017,MignonRisse_2020}. We provide further details on these terms in Section~\ref{sec:sinkcontribution}.

The material coefficients $\kappa_P$, $\kappa_E$ and $\kappa_F$ are the Planck-mean, energy-mean, and flux-mean frequency-integrated specific opacities evaluated in the comoving frame, and are given by,
\begin{gather}
\kappa_{\mathrm{P}} \equiv \frac{\int_{0}^{\infty} \kappa\left(v_{0}\right) B\left(v_{0}, T\right) d v_{0}}{\int_{0}^{\infty} B(\nu_0,T) d\nu_0}, \\
\kappa_{E} \equiv \frac{\int_{0}^{\infty} \kappa\left(v_{0}\right) E_{r0}\left(v_{0}\right) d v_{0}}{\int_{0}^{\infty} E_{r0}(\nu_0)d v_{0}}, \\
\kappa_{F} \equiv \frac{\int_{0}^{\infty} \kappa\left(v_{0}\right) \mathbfit{F}_{r0}\left(v_{0}\right) d v_{0}}{\int_{0}^{\infty} \mathbfit{F}_{r0}(\nu_0)d v_{0}},
\end{gather}
where $B\left(v_{0}, T\right)$ the frequency-dependent Planck function, $E_{r0} (\nu_0)$ the radiation energy density per unit frequency, and $\mathbfit{F}_{r0} (\nu_0)$ the radiation flux per unit frequency, all defined in the comoving frame. The lab-frame and comoving-frame quantities are related by \citep[e.g.,][]{Castor_2004}
\begin{gather}
E_r=E_{r0}+2 \frac{\mathbfit{v} \cdot \mathbfit{F}_{r0}}{c^{2}}+\frac{1}{c^{2}}\left[v^{2} E_{r0}+(\mathbfit{v} \mathbfit{v}): \mathbfss{P}_{r0}\right], \\ 
\mathbfit{F}_r=\mathbfit{F}_{r0}+\mathbfit{v} E_{0}+\mathbfit{v} \cdot \mathbfss{P}_{r0}+\frac{1}{2 c^{2}}\left[v^{2} \mathbfit{F}_{r0}+3 \mathbfit{v}\left(\mathbfit{v} \cdot \mathbfit{F}_{r0}\right)\right], \\ 
\mathbfss{P}_r=\mathbfss{P}_{r0}+\frac{\mathbfit{v} \mathbfit{F}_{r0}+\mathbfit{F}_{r0} \mathbfit{v}}{c^{2}}+\frac{1}{c^{2}}\left[\mathbfit{v} \mathbfit{v} E_{r0}+\mathbfit{v}\left(\mathbfit{v} \cdot \mathbfss{P}_{r0}\right)\right].
\end{gather}
The equations and implementation, by themselves, make no assumptions about the frequency dependence of $\kappa$. Ideally, the correct approach would be to resolve the spectrum of the radiation field, using for example a multigroup method \citep[see, e.g.,][]{Vaytet_2011}, and compute opacities self-consistently. However, this would render the scheme significantly more computationally expensive, and we thus leave it for future extensions. Instead, for the purposes of this work, we shall adopt the approximation that $\kappa_E \approx \kappa_P$ and $\kappa_F \approx \kappa_R$, where $\kappa_R$ is the Rossseland mean opacity given by 
\begin{equation}
    \kappa_{\mathrm{R}}^{-1}=\frac{\int_{0}^{\infty} d \nu_{0} \kappa_{0}\left(\nu_{0}\right)^{-1}\left[\partial B\left(\nu_{0}, T_{0}\right) / \partial T_{0}\right]}{\int_{0}^{\infty} d \nu_{0}\left[\partial B\left(\nu_{0}, T_{0}\right) / \partial T_{0}\right]}.
\end{equation}
The former condition is obtained by assuming the radiation has a blackbody spectrum, and the latter yields the correct radiation force in optically thick media. In Equation~\ref{eq:RTeq} we use $\kappa = \kappa_P$, which would make it consistent with the equation for $E_r$ in steady state. We note that this choice of opacity would not be consistent with the steady state equation for $\mathbfit{F}_r$. However, it is not possible for Equation~\ref{eq:RTeq} to be fully consistent with both the moment equations regardless of the choice of gray opacity $\kappa$ adopted; only a frequency-dependent opacity can permit this.

\subsection{Solution Algorithm}

\subsubsection{Algorithm summary}
\label{sec:overallalgo}

To begin with, it is useful to summarise the series of steps followed by \texttt{VETTAM} in each simulation timestep. We refer the reader to specific subsections for details of each step in the algorithm. 
\begin{enumerate}
    \item Perform the explicit hydrodynamic update (Equations~\ref{eq:explicit}) with the hydrodynamic solver capabilities in \texttt{FLASH}. 
    \item If point sources of radiation are present in the simulation, compute and add their direct contribution to the energy $\left(\dot{E_*}\right)$ and momentum $\left(\dot{\mathbfit{p}_*} \right)$ of the gas (Section~\ref{sec:sinkcontribution}). 
    \item Use the gas variables to compute opacities ($\kappa_P$, $\kappa_R$) and the source function ($S$) for the transfer equation. Solve the time-independent transfer equation using the hybrid characteristics ray-tracer and compute the Eddington Tensor from the solution $\mathbfss{T}$ (Section~\ref{sec:ComputeVET}). 
    \item Perform a linearised first-order backwards Euler implicit update for the equations governing $E_r$ and $\mathbfit{F}_r$ with the temperature obtained from step~ii, keeping the hydrodynamic quantities fixed for this update (Section~\ref{sec:PicardUpdate}). This update, converged to a relative tolerance of $\epsilon_{\mathrm{R}}$, provides a guess solution for the radiation quantities $E_{r,*}$ and $\mathbfit{F}_{r,*}$.
    \item Solve the nonlinear equation for a guess for the gas temperature $T_*$ with a Newton's method to a relative tolerance of $\epsilon_{\mathrm{N}}$ (Section~\ref{sec:PicardUpdate}). 
    \item Repeat steps~iv \&~v until the vector of quantities $\mathbfit{x}_{\mathrm{P}} = \left[E_{r,*},\mathbfit{F}_{r,*},T_* \right]$ converges to a relative tolerance of $\epsilon_{\mathrm{P}}$. Set the time-updated values for these variables to the converged guess.
    \item Add the explicitly-handled radiation source terms $cG_{0,e}$ (Equation~\ref{eq:G0explicit}) and $\mathbfit{G}$  (Equation~\ref{eq:G}) using the converged solution for $\mathbfit{x}_{\mathrm{P}}$. 
    \item Update the time $t^{n+1} = t^{n} + \Delta t$, calculating the new timestep according to a modified Courant-Friedrichs-Lewy (CFL) condition using the adiabatic sound speed for RHD.
\end{enumerate}

The modified timestep here is essentially a modification to the standard CFL condition \citep{Courant_1928}, modified to account for the effect of radiation pressure on the propagation of acoustic waves \citep{Mihalas_1984}. Following \citet{Krumholz_2007a}, we use an approximate expression for the effective sound speed,
\begin{equation}
c_{\rm eff} = \sqrt{\frac{\gamma P + (4/9) E_r(1-e^{-\rho\kappa_{R}\Delta x})}{\rho}},
\end{equation}
where $\gamma$ is the adiabatic index of the gas, and $\Delta x$ is the computational cell length on the highest refined level $l_{\rm max}$, as is the convention in FLASH. The hydrodynamic timestep is then set to
\begin{equation}
\label{eq:timestep}
\Delta t = C_{0} \frac{\Delta x}{\mathrm{max}(|\mathbfit{v}|+c_{\rm eff})}, 
\end{equation}
where $C_0$ is the Courant number, and the denominator denotes the maximum signal speed at $l_{\rm max}$.

\subsubsection{Operator-Splitting}
\label{sec:opsplitting}
Equations~\ref{eq:continuityeq} --~\ref{eq:Frad} are a set of coupled, nonlinear, hyperbolic conservation laws plus source terms, for which various methods exist to obtain solutions. However, the large difference in hydrodynamical sound-crossing and the radiation light-crossing timescales poses a difficult numerical challenge. In addition, the stiff nonlinear source terms associated with the radiation-gas interaction could render the system sensitive to perturbations and prone to ringing \citep{Leveque_2002}. Thus, we must solve the radiation subsystem along with the coupled stiff source term update for the hydrodynamic quantities in an implicit manner. For this purpose, we operator-split this subset of equations from the hyperbolic hydrodynamic update that contains non-stiff source terms, which is treated explicitly using the preexisting infrastructure available in \texttt{FLASH}. We also treat the contribution of radiation source terms in the gas momentum density equation ($\mathbfit{G}$ in Eq.~\ref{eq:gasMom}) explicitly. In the gas energy equation, we treat the update for the stiff gas-radiation interaction term implicitly and by default we treat the other $\mathcal{O}(v/c)$ terms explicitly. However, in some cases where the system is in the dynamic diffusion regime, we found treating these terms implicitly as well rendered the system more robust and stable at larger timesteps, and our software implementation therefore provides a runtime switch to specify whether to treat the non-stiff energy source terms implicitly or explicitly. 

Formally, we can express our operator splitting approach in terms of the following sub-problems:
\begin{eqnarray}
\label{eq:explicit}
\frac{\partial \mathbfit{U}_e}{\partial t} + \nabla \cdot (\mathbb{F}_e) &=& \mathbfit{S}_e,\\
\label{eq:implicit}
\frac{\partial \mathbfit{U}_r}{\partial t} + \nabla \cdot (\mathbb{F}_r) &=& \mathbfit{S}_r,
\end{eqnarray} 
where
\begin{eqnarray}
\label{eq:explicit1}
	\mathbfit{U}_e &=& 
	\begin{bmatrix}
		 \rho \\ \rho \mathbfit{v} \\ E 
	\end{bmatrix}
	,\\
\label{eq:explicit2}
	\mathbb{F}_e &=& 
	\begin{bmatrix}
		\rho \mathbfit{v}  \\ \rho \mathbfit{v}\mathbfit{v} + P \mathbfss{I}  \\ (E+P)\mathbfit{v}
	\end{bmatrix}
	,\\
\label{eq:explicit3}
	\mathbfit{S}_e &=& 
	\begin{bmatrix}
		0 \\ -\rho \nabla \mathbf{\Phi} + \mathbfit{G}\\ -\rho \textbf{v} \cdot \nabla \mathbf{\Phi} + cG_{0,e}
	\end{bmatrix}
	,\\
\label{eq:implicit1}
	\mathbfit{U}_r &=& 
	\begin{bmatrix}
		E \\ E_R \\ \mathbfit{F}_r 
	\end{bmatrix}
	,\\
\label{eq:implicit2}
	\mathbb{F}_r &=& 
	\begin{bmatrix}
		0 \\ \mathbfit{F}_r \\ c^2E_R \mathbfss{T}
	\end{bmatrix}
	,\\
\label{eq:implicit3}
	\mathbfit{S}_r &=& 
	\begin{bmatrix}
		 -\rho \kappa_Pc [a_RT^4-E_R] \\ -cG^0 \\ -c^2 \mathbfit{G}
	\end{bmatrix}.
\end{eqnarray}
In the expressions above,
\begin{equation}
    \label{eq:G0explicit}
    \begin{aligned}
    cG_{0,e} =& \rho \left(\kappa_F - 2\kappa_E \right) \frac{\mathbfit{v} \cdot \mathbfit{F}_r}{c^2} \\ &+ \rho \left( \kappa_E - \kappa_F \right) \left[\frac{v^2}{c^2}E_r + \frac{\mathbfit{v}\mathbfit{v}}{c^2} :\mathbfss{P}_r  \right]
    \end{aligned}
\end{equation}
is the collection of $\mathcal{O}(v/c)$ coupling terms that by default we treat explicitly; in the alternative implicit treatment we move this term from the last element of $\mathbfit{S}_e$ to the first element of $\mathbfit{S}_r$.

We solve subsystem~\ref{eq:explicit} using the pre-existing infrastructure available in \texttt{FLASH}, with the trivial modification of adding the radiation-related source terms $\mathbfit{G}$ and $cG_{0,e}$ that appear in $\mathbfit{S}_e$. We discretise these terms such that $E_r$ and $\mathbfit{F}_r$ are the values obtained after the implicit update for the radiation quantities, i.e., at time $t^{n+1}$.

On the other hand, subsystem~\ref{eq:implicit} represents a set of coupled nonlinear equations that we update fully implicitly using a first-order backward Euler differencing in time. We restrict the temporal accuracy to first-order because higher-order implicit time integration schemes have been found to lead to oscillatory solutions when using large time steps \citep{Sekora_2010}. We use a Godunov method to discretise the vector flux $\nabla \cdot \mathbb{F}_r$ for the conserved quantities $\mathbfit{U}_r$, using an HLLE Riemann solver in an implicit fashion by defining the variables in the flux expression to be at time $t^{n+1}$. We describe this procedure in further detail in Section~\ref{sec:hllefluxes}. During this stage, we keep the hydrodynamical quantities $\rho$ and $\mathbfit{v}$ fixed to the state obtained after the hydrodynamic update in the source terms on the right hand side of Equation~\ref{eq:implicit}. On the other hand, the radiation quantities $E_r$ and $\mathbfit{F}_r$ are at time $t^{n+1}$, as required for an implicit method.

The above discretisation approach, when written down for every cell in the domain, leads to a system of nonlinear equations that can be represented in a matrix form, and inverted to obtain a solution. We use a fixed-point Picard iteration scheme to treat the nonlinear update of the implicit subsystem and describe the method in further detail in Section~\ref{sec:PicardUpdate} below. We also note that the Eddington Tensor ($\mathbfss{T}$), obtained with a solution to the time-independent radiative transfer equation, is pre-computed using the physical quantities obtained after the hydrodynamic update, and is kept fixed for the implicit update. In Sections~\ref{sec:ComputeVET} and \ref{sec:sinkcontribution} we provide further details on how $\mathbfss{T}$ is obtained with the hybrid characteristics ray-tracing scheme for diffuse sources and for point sources of radiation, respectively.

\subsubsection{Implicit Hyperbolic Transport of Radiation}
\label{sec:hllefluxes}

To evolve the hyperbolic transport equations for $E_r$ and $\mathbfit{F}_r$ in~\ref{eq:implicit}, we use a first-order Godunov finite-volume method using a Harten-Lax-van Leer (HLL)-type Riemann solver \citep{Toro_1997} to compute the flux of the conserved variables. With this approach, similar to the one described by \citet{Jiang_2012}, the discretised evolution equation can be written as 
\begin{equation}
\begin{aligned}
(\mathbfit{U}_r)_{i,j,k}^{n+1} = &(\mathbfit{U}_r)_{i,j,k}^{n} - \frac{\Delta t}{\Delta x} \left[\mathbfit{F}_{i+1/2,j,k}^{\mathrm{HLLE}} - \mathbfit{F}_{i-1/2,j,k}^{\mathrm{HLLE}} \right] \\ &- \frac{\Delta t}{\Delta y} \left[\mathbfit{F}_{i,j+1/2,k}^{\mathrm{HLLE}} - \mathbfit{F}_{i,j-1/2,k}^{\mathrm{HLLE}} \right] \\ &- \frac{\Delta t}{\Delta z} \left[\mathbfit{F}_{i,j,k+1/2}^{\mathrm{HLLE}} - \mathbfit{F}_{i,j,k-1/2}^{\mathrm{HLLE}} \right] + \Delta t \mathbfit{S}_r^{n+1},
\end{aligned}
\end{equation}
where the terms $\mathbfit{F}^{\mathrm{HLLE}}$ are the vector of fluxes for the conserved quantities at each cell interface computed by an HLLE Riemann solver (Equation~39 of \citealt{Sekora_2010}). The left and right states at the interface for the Riemann solver are obtained using a piecewise constant (first-order) reconstruction, using the state of the conserved quantities $\mathbfit{U}_r$ at time $t^{n+1}$ when computing the HLLE fluxes. The characteristic left/right going wavespeeds ($C_{\mathrm{HLLE}}^{\mathrm{L}}$ and $C_{\mathrm{HLLE}}^{\mathrm{R}}$ respectively) are given by 
\begin{equation}
C_{\mathrm{HLLE}} =\sqrt{f}c \sqrt{\frac{1-e^{-\tau_c} }{\tau_c}},
\end{equation}
where 
\begin{equation}
    \label{eq:taucorrect}
    \tau_c \equiv\left(10\,\Delta\ell\, \rho\,\kappa_{R}\right)^{2} /(2 f),
\end{equation}
$\Delta \ell$ is the cell thickness and $f$ is the diagonal component of $\mathbfss{T}$ in the direction of the flux. We obtain this relation by using the eigenvalues of the radiation moment equations in the free streaming limit ($\pm \sqrt{f}c$), and applying an optical depth-dependent correction factor following Equation A3 of \citet{Jiang_2013} (\citetalias{Jiang_2013} hereafter) to circumvent the issue of the numerical diffusive flux becoming dominant over the physical diffusion flux in the optically thick regime \citep{Audit_2002}. In Appendix~\ref{sec:wavespeed} we show that this correction is required to avoid substantial numerical diffusion in simulations where the cell optical depth is $\gg 1$.

However, in our implementation, we introduce a modification to the \citetalias{Jiang_2013} correction factor for AMR grids. \citetalias{Jiang_2013} evaluate the correction factor $\tau_c$ at a cell interface using the arithmetic mean of the values computed using cell-centred quantities at the left ($\tau_\mathrm{L}$) and right ($\tau_\mathrm{R}$) cells of the interface. We do the same at all interfaces \textit{except} at AMR level boundaries, where we use the upstream value to avoid biases arising from the different cell sizes on the two sides of the interface. We found this modification to be necessary for obtaining continuous and accurate results with AMR\footnote{The implementation of \citetalias{Jiang_2013} was for uniform grids, and hence did not face the aforementioned issue.}, and justify this choice in Appendix~\ref{sec:wavespeedamr}. In summary, the value of $\tau_c$ at an interface for the right- ($C_{\mathrm{HLLE}}^{\mathrm{R}}$) and left- ($C_{\mathrm{HLLE}}^{\mathrm{L}}$) going waves is
\begin{equation}
    \label{eq:wavespeedtau}
    \tau_c(\tau_\mathrm{L},\tau_\mathrm{R}) = 
    \begin{cases}
    \frac{\tau_\mathrm{L}+\tau_\mathrm{R}}{2} \; &|l_\mathrm{R} - l_{\mathrm{L}}| = 0 \\
    \tau_L \; &|l_\mathrm{R} - l_{\mathrm{L}}| > 0 \; \text{for } C_{\mathrm{HLLE}}^{\mathrm{R}} \\
    \tau_R \; &|l_\mathrm{R} - l_{\mathrm{L}}| > 0 \; \text{for } C_{\mathrm{HLLE}}^{\mathrm{L}}
    \end{cases},
\end{equation}
where $l_\mathrm{R}$ and $l_\mathrm{L}$ are the numbers of the AMR levels of the cells to the right and left of the interface, respectively. 

We must also take steps to ensure that the hyperbolic transport of the conserved quantities $E_r$ and $\mathbfit{F}_r$ retains conservation at cell interfaces where there is a jump in refinement level. For instance, the net transport of radiation energy and momentum out of a coarse cell should be balanced by the corresponding sum of the same entering the finer neighbour cells. However, this condition is not automatically satisfied at AMR level boundaries, since the terms that express the transport fluxes depend on ghost cell data at block boundaries, which are interpolated, and are thus not identical at both sides of the interface. In an explicit method, it is possible to perform a correction step after the hyperbolic update is performed to ensure flux consistency \citep{Berger_1989}. However, this luxury is not available to an implicit method, and flux consistency should either by enforced by construction in the implicit set of equations solved \citep[e.g.,][]{Commercon_2011,Klassen_2014}, or through a level-by-level approach with synchronisation steps that also allows one to use adaptive time-stepping \citep[e.g.,][]{Howell_2003,Zhang_2011,Commercon_2014}. Since \texttt{FLASH} does not include adaptive time-stepping, we chose to adopt the former approach. Specifically, we enforce conservation by replacing the coarse flux determined by the HLLE solver across any coarse-fine interface with the sum of the fine fluxes. For example, consider the case of a 2D coarse-fine interface, and denote the hyperbolic flux out of the coarse cell as $F^{\mathrm{HLLE}}_\mathrm{c}$, and that entering the two fine cells to be $F^{\mathrm{HLLE}}_\mathrm{f1}$ and $F^{\mathrm{HLLE}}_\mathrm{f2}$. Since the fine cells each have face areas equal to half that of the coarse cell, exact conservation requires that
\begin{equation}
    F^{\mathrm{HLLE}}_\mathrm{c} = \frac{1}{2}\left(F^{\mathrm{HLLE}}_\mathrm{f1} + F^{\mathrm{HLLE}}_\mathrm{f2}\right).
    \label{eq:interface_flux}
\end{equation}
While in general the HLLE solver will not enforce exact equality, we do in our scheme by explicitly replacing $F^{\mathrm{HLLE}}_\mathrm{c}$ with the right hand side of Equation~\ref{eq:interface_flux} when writing out the discretised equation to be solved.  This ensures that our scheme achieves conservation by construction.

\subsubsection{Implicit Nonlinear Update}
\label{sec:PicardUpdate}

The system of equations described by the subsystem~\ref{eq:implicit} represent a nonlinear coupled set of equations, where the non-linearity arises due to the stiff gas-radiation interaction term and the nonlinear nature of the temperature dependence of $\kappa_P$ and $\kappa_R$ \citep[e.g.,][]{Semenov_2003}. There exist numerous strategies for solving systems of coupled nonlinear equations \citep[see, e.g.,][]{Kelley_1995}, and our choice is dictated by simplicity and performance. Most commonly, Newton-Raphson iteration methods are used to treat such systems; however, they require the computation of the Jacobian of the system, which in our case is unavailable analytically, and would be expensive to compute numerically. While there exist Jacobian-free Newton-Krylov methods to circumvent costs associated with constructing the Jacobian \citep{Knoll_2004}, we adopt a simpler, yet robust, fixed-point Picard iteration scheme\footnote{This iterative method is also sometimes called nonlinear Richardson iteration, or the method of successive substitution.}. Picard iteration is a method for solving a system of nonlinear equations by reformulating them as the problem of finding the fixed point of a function \citep{Kelley_1995,Burden_1997}. This is done by starting with an initial guess for the solution to the nonlinear system, and successively improving the guess through the solution of a simpler linearised recasting of the nonlinear system of equations. 

We implement the Picard iteration method in our scheme in the following fashion: we first operator-split the gas energy update from the radiation moment equations, and then discretise the term proportional to $T^4$ in the latter to use a provided guess temperature $T_*$, which we set to the old time value ($T_{n}$) at the start of the update. The first-order Euler backward update for a timestep $\Delta t = t_{n+1} - t_n$ can be written for each computational cell as
\begin{gather}
    \label{eq:PicardEr}
    \frac{E_{r,*} - E_{r,n}}{\Delta t} + \nabla \cdot \mathbfit{F}_{r,*} = cG^{0}_*  \\
    \frac{\mathbfit{F}_{r,*} - \mathbfit{F}_{r,n}}{\Delta t} + c^2\nabla \cdot \mathbfss{T}E_{r,*} = c\mathbfit{G}_*,
    \label{eq:PicardFr}
\end{gather}
where the source terms $cG^{0}_*$ and $\mathbfit{G}_*$ use the guess temperature $T_*$, and corresponding opacities $\kappa_P(T_*), \kappa_R(T_*)$  in their expressions. The discretisation described here effectively linearises the implicit radiation moment equation update, and we use sparse matrix solvers based on Krylov subspace methods \citep{Saad_2003} offered by the \texttt{PETSc} library to obtain the solution to $E_{r,*}$ and $\mathbfit{F}_{r,*}$. We use the generalised minimum residual (GMRES) solver \citep{Saad_1986} by default, but allow users to choose other solvers and preconditioners at runtime. We use the default convergence criteria in \texttt{PETSc} based on the 2-norm of the preconditioned residual to check for convergence to a user-defined relative tolerance $\epsilon_{\mathrm{R}}$. The guess for the gas temperature is then improved by implicitly updating the gas energy as
\begin{equation}
    \label{eq:PicardTemp}
    \frac{E_{*} - E_{n}}{\Delta t} = -\rho \kappa_P c \left[a_R T_*^4 - E_{r,*} \right],
\end{equation}
where $\kappa_P$ is identical to that used in Equations~\ref{eq:PicardEr} and \ref{eq:PicardFr}, and $E_{*} - E_{n} = \alpha(T_{*} - T_{n})$, using the relation between internal energy and temperature $E_{\mathrm{int}} = \alpha T$, and the fact that the discretisation we use ensures the kinetic energies cancel out. This represents a nonlinear equation for the new temperature guess $T_*$, which is, however, local, and hence we can solve this independently for each cell. We use a simple Newton's method by analytically constructing the Jacobian of the polynomial equation~\ref{eq:PicardTemp} to obtain $T_*$, assessing convergence on the relative tolerance of the temperature $\epsilon_{\mathrm{N}}$. The combination of the implicit radiation subsystem update, and the implicit gas temperature update comprise \textit{one} Picard iteration. At the end of each iteration, we check for the residual change in the vector
\begin{equation}
    \label{eq:Picardvec}
    \mathbfit{x}_{\mathrm{P}} = 
    \begin{bmatrix}
		 E_r \\ \mathbfit{F}_r \\ T
	\end{bmatrix},
\end{equation}
over the iteration, and check for convergence within a relative tolerance $\epsilon_{\mathrm{P}}$. If convergence is satisfied, we set the new time solution for $E_r$, $\mathbfit{F}_r$ and $T$ to be equal to the guess in the last Picard iteration, and if not we repeat the procedure, using the values obtained at the end of this iteration as our new guess.

One may notice that the radiation-gas interaction terms in Equations~\ref{eq:PicardEr} and \ref{eq:PicardTemp} are not the same by construction, with the former using a guess temperature $T_*$ in $cG^0_*$ that is lagged by one Picard iteration as compared to the $T_*$ used in the latter. This, in some cases, can lead to non-conservation of total energy in the domain. We thus add a correction term $\Delta E$ explicitly in each Picard iteration to the right-hand side of Equation~\ref{eq:PicardTemp} accounting for this variation in discretisation. This has the form
\begin{equation}
    \Delta E = \rho \kappa_Pc a_{R}\left[T_{n,*}^4 - T_{o,*}^4 \right],
\end{equation}
where $T_{o,*}$ is the value of $T_*$ used in the last update of Equation~\ref{eq:PicardEr}, and $T_{n,*}$ is that used in the last update of Equation~\ref{eq:PicardTemp}. This ensures that energy is conserved irrespective of the adopted value of $\epsilon_\mathrm{P}$. However, we note that since we use implicit updates for Equations~\ref{eq:PicardEr} and \ref{eq:PicardTemp}, we cannot ensure strict convergence of energy to machine precision, and are limited to the precision $\epsilon_{\mathrm{N}}$ and $\epsilon_{\mathrm{R}}$. 

\subsubsection{Computing the Eddington Tensor}
\label{sec:ComputeVET}

The VET, used to close the radiation momentum equations with Equation~\ref{eq:closure}, is calculated explicitly from a formal solution of the time-independent radiative transfer equation (Equation~\ref{eq:RTeq}). We use a hybrid-characteristics based raytracing approach to solve this equation on large sets of characteristics (rays) using the implementation in \texttt{FLASH} by \citet{Buntemeyer_2016}. We use the obtained solution for the gray radiation intensity ($I_r$) from the ray-tracer, perform angular quadratures on it to compute $E_r$ and $\mathbfss{P}_r$, and use them to obtain $\mathbfss{T}$ using Equation~\ref{eq:closure}. To avoid having to allocate and store the specific intensity over all angles and spatial locations, we compute the quadratures on-the-fly for the intensity along each ray in space. The discretised quadrature contributions are given by
\begin{equation}
E_r=\frac{c}{4\pi}\sum_{k=0}^{N_{\Omega}-1} w_{k} I_{r,k} , 
\end{equation}
and 
\begin{equation}
\mathbfss{P}_r= \frac{c}{4 \pi}\sum_{k=0}^{N_{\Omega}-1} w_{k} I_{r,k} \mu_{i k} \mu_{j k} ,
\end{equation}
where $I_{r,k}$ is the intensity along a ray in the direction $\hat{\mathbfit{n}}_k$, $w_k$ is the quadrature weight, $\mu_{ik}$ = $\hat{\mathbfit{n}}_k \cdot \hat{\mathbfit{x}}_i$ where $\hat{\mathbfit{x}}_i$ is the unit vector along the coordinate axis $i$, and the quadrature sum is performed over $N_{\Omega}$ discrete angles. We use the \texttt{HEALPIX} tesselation scheme to discretise angles on the unit sphere uniformly, which allows values of $N_{\Omega} = 12 N_{\mathrm{side}}^2$ where $N_{\mathrm{side}}$ is an integer that is a power of 2 (i.e., $N_{\mathrm{side}} = 1,2,4,8,...$). We expect the appropriate value of $N_{\Omega}$ to use to be problem-dependent; however, we find reasonable results for our tests even with moderate $N_{\Omega}$, as shown in Appendix~\ref{sec:angularres}. We also randomly rotate the angles generated by the \texttt{HEALPIX} tesselation to prevent accumulation of artefacts introduced by the discretisation \citep[see, e.g.,][]{Krumholz_2007}. We  note that the VET is computed at the start of the time step, and kept fixed for the overall radiation system update described in Section~\ref{sec:PicardUpdate}.

\subsubsection{Point Sources Contribution}
\label{sec:sinkcontribution}

\texttt{VETTAM} is a hybrid radiation transport scheme, i.e., it splits the radiation field into a direct and diffuse component \citep{Wolfire_1986,Murray_1994}, where the direct component includes the contributions from point sources implemented with sink particles \citep{Federrath_2010_Sinks}, and the diffuse component involves the diffuse (re-)emission of (thermal) radiation by the dust. The latter is handled by the radiation moment equations closed with the VET described above. The direct contribution is handled solely by the hybrid-characteristics ray tracer, which solves Equation~\ref{eq:RTeq} along rays that originate at point sources in the domain, with no effective emission, i.e., $S=0$. This splitting is useful if a simulation includes sink particles to represent stars or clusters, whose contribution to the radiation field can be quite asymmetric depending on the matter distribution. This splitting approach also allows a frequency-dependent treatment of the direct radiation \citep[e.g.,][]{Kuiper_2010,Rosen_2016}, which often has a very different colour temperature than the reprocessed radiation, and thus experiences very different matter opacities \citep[see][for a direct comparison]{Kuiper_2012}. The ray-trace is performed using the implementation originally described in \citet{Rijkhorst_2006}, and improved later by \citet{Peters_2010} and \citet{Buntemeyer_2016}. The ray-tracer computes effective optical depths from a point source to each cell in the domain and used to obtain the energy ($\dot{E_*}$) and momentum ($\dot{\mathbfit{p}_*}$) deposited in the gas. The energy deposition rate $\dot{E_*}$ absorbed by the gas in a computational cell at a distance $r$ from the star is given by 
\begin{equation}
    \dot{E_*} = \frac{L_* e^{-\tau_*} \left( 1-e^{-\tau_{\mathrm{cell}}} \right)}{4 \pi r^2 \Delta r},
\end{equation}
where $L_*$ is the luminosity of the point source, $\tau_*$ is the optical depth to the cell for a ray originating at the point source, $\tau_{\mathrm{cell}} = \rho \kappa_* \Delta r$ is the local optical depth of the cell where $\kappa_*$ is the opacity to the direct radiation, and $\Delta r$ is the length of the ray intersected by the cell. For numerical stability, when $\tau_{\mathrm{cell}}$ is very small, $\dot{E_*}$ is estimated by a Taylor-expanded form of the above relation given by
\begin{equation}
    \dot{E_*} =  \frac{\tau_{\mathrm{cell}} L_* e^{-\tau_*}}{4 \pi r^2 \Delta r}.
\end{equation}
The momentum contribution rate $\dot{\mathbfit{p}_*}$ is given by 
\begin{equation}
    \dot{\mathbfit{p}_*} = \frac{\dot{E_*}}{c} \hat{\mathbfit{r}},
\end{equation}
where $\hat{\mathbfit{r}}$ denotes the direction of the ray to the cell from the point source.

\section{Numerical Tests}
\label{sec:Tests}
In this section we provide numerical tests of the scheme described in the previous sections. We compare our numerical results ($f_{\mathrm{num}}$) with analytic or semi-analytic solutions ($f_{\mathrm{an}}$) when available, using either the $L_1$ relative norm or the maximum relative error $L_{\mathrm{max}}$ defined by 
\begin{equation}
    L_1 = \frac{\sum_i|f_{\mathrm{num,i}} - f_{\mathrm{an,i}}| \,\Delta x_i}{\sum_if_{\mathrm{an,i} \, \Delta x_i}},
\end{equation}
and 
\begin{equation}
    L_{\mathrm{max}} = \max_i\left[ {\frac{|f_{\mathrm{num,i}} - f_{\mathrm{an,i}}|}{f_{\mathrm{an,i}}}}\right], 
\end{equation}
where $i$ can denote the solution at a spatial location or time $t_i$ for the problem. We use the following settings for our tests, unless otherwise specified: relative tolerances of $\epsilon_{\mathrm{N}} = \epsilon_{\mathrm{R}} = 10^{-6}$ and $\epsilon_{\mathrm{P}} = 10^{-3}$, GMRES solver left-preconditioned with the additive Schwarz method (ASM) for the implicit radiation update, and a Courant number $C_0 = 0.8$. We also do not use gravity in any of our tests (i.e. $\phi = 0$).

\subsection{Radiating Pulse}
\label{sec:Radiating_Pulse}
In our first test, we evolve the propagation of a one-dimensional Gaussian pulse of radiation energy in a static medium ($v = 0$), for 3 different opacities ($\kappa_0$), that correspond to the streaming ($\tau \ll 1$), weak equilibrium diffusion ($\tau \sim 1$) and strong equilibrium diffusion ($\tau \gg 1$) regimes respectively. In the streaming regime, radiation and hydrodynamics are decoupled and the resulting dynamics resemble an advection process. In the diffusion limits, radiation and hydrodynamics are strongly coupled, and the resulting dynamics resemble a diffusion process. We perform this test to demonstrate that our scheme is capable of reproducing the right solutions in all regimes of radiation transport. 
The test setup is a 1D domain ranging from $x = -x_0$ to $x = x_0$, where $x_0 = 0.5$ cm for the streaming test, and $x_0 = 5$ cm for the diffusion tests respectively; the domain size is larger in the diffusion tests so that we can capture the diffusion of the pulse for longer times without boundary effects coming into play. The density of the gas for all cases is fixed to $\rho = 10^{-20} \, \mathrm{g} \,\mathrm{cm}^{-3}$, and we disable hydrodynamics, so $\rho$ does not evolve and the gas velocity remains $v=0$. The initial radiation energy density is
\begin{equation}
\label{eq:PulseDiffusionEr}
    E_r (x,0) = E_0\exp \left(-\mu^2 x^2  \right), 
\end{equation}
where we set $E_0 = 1$ erg cm$^{-3}$ and $\mu = 20$ cm$^{-1}$. Since the purpose of this test is to check whether our scheme captures the physical transport/diffusive fluxes accurately, we set $\kappa_E = 0$ and $\kappa_F = \kappa_0 = 1, 4 \times 10^{21}$, and $4\times 10^{24}$ cm$^2$ g$^{-1}$ for the streaming, static diffusion and equilibrium diffusion versions respectively; this has the effect of disabling energy exchange between gas and radiation (since all exchange terms are proportional to either $\kappa_E$ or $v$, both of which are zero), and thus mimics the effects of a purely scattering medium. For the streaming test, we initialise the radiation flux to the streaming solution,
\begin{equation}
    F_r (x,0) = cE_r(x,0) ,
\end{equation}
whereas we use the solution expected for pure diffusion for diffusion tests, i.e., 
\begin{equation}
\label{eq:PulseDiffusionFr}
    F_r(x,0) = \frac{-c}{3\rho \kappa_0}\frac{\partial E_r(x,0)}{\partial x} = \frac{2c \mu^2 x^2}{3} E_r(x,0).
\end{equation}
We do not use the ray-tracer for this test as the Eddington tensor component ($f_{xx}$) is spatially and temporally uniform with a value of $f_{xx} = 1$ for the streaming test and $f_{xx} = 1/3$ for the others. We evolve the system at the light crossing timescale across a cell, i.e., $\Delta t = \Delta x/c$, where $\Delta x$ is the cell thickness. We use a resolution of 1024 cells for our tests, and adopt periodic (outflow) boundary conditions for the streaming (diffusion) tests. In the streaming limit, the exact solution is a radiation energy density profile identical to the initial state, displaced by $ct$ in the direction of the initial flux, i.e., 
\begin{equation}
    \label{eq:Streaming_solution}
    E_r(x,t) = E_r(x-ct,0).
\end{equation}
In the diffusion tests, one can obtain an analytic solution by the method of Green's functions, which gives
\begin{equation}
    \label{eq:Diffusion_solution}
    E_r(x,t) = \frac{1}{\left(4 D t \mu^{2}+1\right)^{1 / 2}} \exp \left(\frac{-\mu^2x^{2}}{4 D t \mu^{2}+1}\right).
\end{equation}
The corresponding flux is $F_r(x,t) = -D (\partial E_r(x,t)/ \partial x)$, where $D=c/3\kappa_0\rho$ is the diffusion coefficient.

\begin{figure*}
    \centering
    \includegraphics[width=0.98\textwidth]{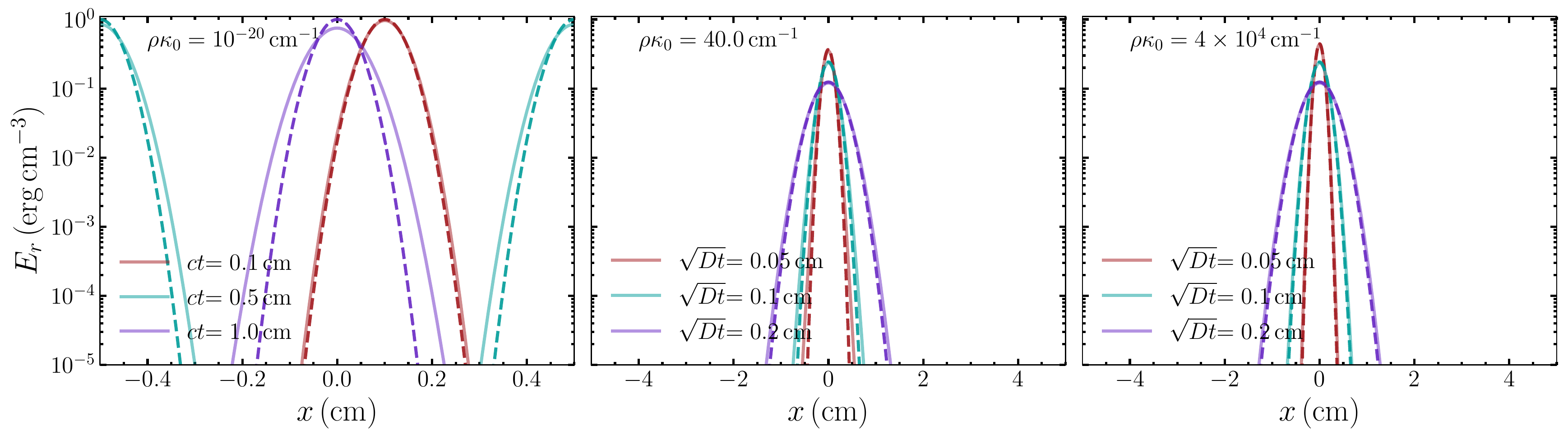}
    \caption{Radiation pulse in the streaming (left), weak equilibrium diffusion (middle) and strong equilibrium diffusion (right) regimes for three different times, with their corresponding analytical solutions given by Equation~\ref{eq:Streaming_solution} (streaming) and Equation~\ref{eq:Diffusion_solution} (weak/strong diffusion) overplotted (dashed lines). We indicate the total opacity ($\rho \kappa_0$) for each version in the top-left corner of the plot.}
    \label{fig:pulse_test}
\end{figure*}

We compare the exact and numerical results we obtain for the radiation energy ($E_r$) for three times in all three cases in Figure~\ref{fig:pulse_test}. We see that the agreement with the analytical solutions is good, especially for the diffusing pulses. The $L_1$ relative errors for the three time instances shown in Figure~\ref{fig:pulse_test} for the static diffusion test are 5\%, 3.7\% and 3.3\%, and for the equilibrium diffusion test are 2.4\%, 3\% and 3.2\%. The agreement is poorer for the streaming pulse though, especially at later times, with  $L_1$ relative errors of 3.7\%, 16\% and 27\% respectively. Although we capture the propogation speed (hence, position) of the pulse accurately in the streaming regime, we find that the pulse has diffused from its initial true state due to numerical diffusion at later times. This is not surprising considering that our scheme is only first-order accurate in space and time. In addition, it is widely known that implicit methods perform poorly when trying to capture the propagation of individual wave modes \citep{Sekora_2010}. We keep these limitations in mind, and aim to address this with higher order reconstruction strategies in future versions of the code. 

\subsection{Dynamic Diffusion Test}
The dynamic diffusion regime ($\tau \beta\gg 1$) is a regime of high optical depth where the photons are effectively trapped in the fluid so strongly that radiation transport is primarily by the advection of photons by the gas, rather than diffusion of photons through the gas. Reproducing this limit of radiation hydrodynamics requires accurate handling of the $\mathcal{O}(v/c)$ source terms in the radiation moment equations. To test whether our scheme can achieve this, we setup a Gaussian pulse test with a domain and initial conditions similar to those used in diffusion tests in the previous section, including the condition of $\kappa_E = 0$. However, we increase the value of $\kappa_F = \kappa_0 = 4 \times 10^{26}$ cm$^2$ g$^{-1}$, and initialise the gas with a velocity $v = 3 \, \mathrm{km} \, \mathrm{s}^{-1}$ in the positive $x$ direction rather than 0. The Eddington approximation is used to estimate the Eddington tensor. In addition, we also take care to modify the initial values of $E_r$ and $F_r$, which are defined in the lab frame in our scheme, by performing the appropriate Lorentz transformation from the comoving frame initial conditions (given by Equations \ref{eq:PulseDiffusionEr} and \ref{eq:PulseDiffusionFr}) to the lab frame \citep{Mihalas_1984}. The domain is discretised with 2048 uniformly spaced cells. 

\begin{figure}
    \centering
    \includegraphics[width = \columnwidth]{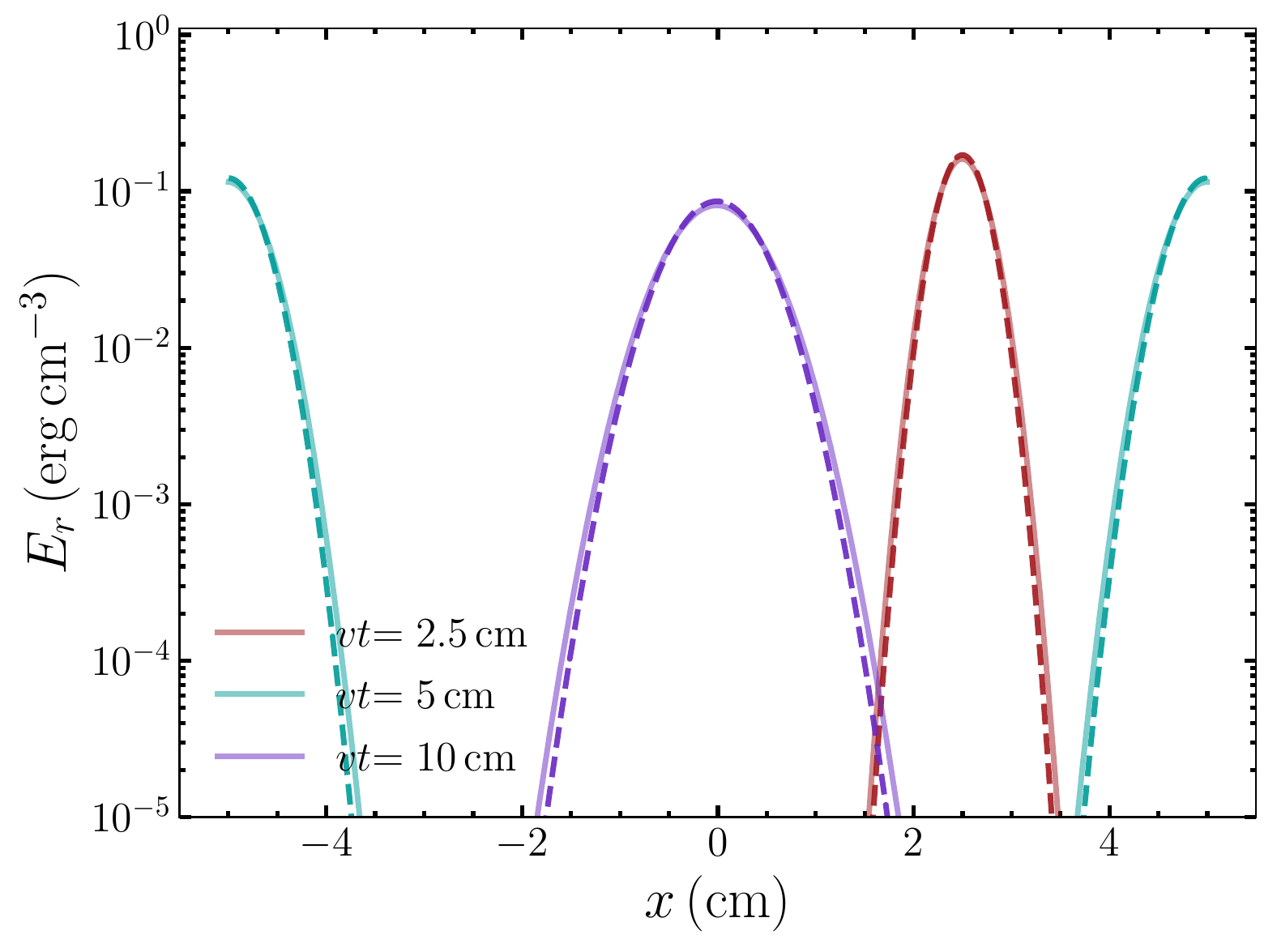}
    \caption{Lab-frame radiation energy density ($E_r$) for a Gaussian radiation pulse in the dynamic diffusion regime, with $\rho \kappa_0 = 4 \times 10^{6}$ cm$^{-1}$ and $v/c \sim 10^{-5}$, for three different times corresponding to 25\%, 50\% and 100\% of the domain crossing time. Solid lines show the numerical solution we obtain, while dashed lines indicate the corresponding analytical solutions.}
    \label{fig:DynamicDiffusion}
\end{figure}

We show our numerical results in Figure~\ref{fig:DynamicDiffusion} for times corresponding to 25\%, 50\% and 100\% of the domain crossing time. We compare this solution with the expected analytical solution for this system, which in the comoving frame should be identical to Equation~\ref{eq:Diffusion_solution}. In the simulation frame, the corresponding solution is
\begin{equation}
    E_r(x,t) = \frac{1}{\left(4 D t \mu^{2}+1\right)^{1 / 2}} \exp \left(\frac{-(\mu(x-vt))^{2}}{4 D t \mu^{2}+1}\right).
\end{equation}
We find that our numerical solution is in good agreement with the analytical one, with $L_1$ relative errors of 4.5\%, 5.6\% and 5.8\% respectively for the three timestamps shown in Figure~\ref{fig:DynamicDiffusion}. This test demonstrates the capability of our scheme to perform correctly in the dynamic diffusion regime.  

\subsection{Non-Equilibrium Radiation-Matter Coupling Test}
\label{sec:RadEqTest}
\begin{figure}
    \centering
    \includegraphics[width=\columnwidth]{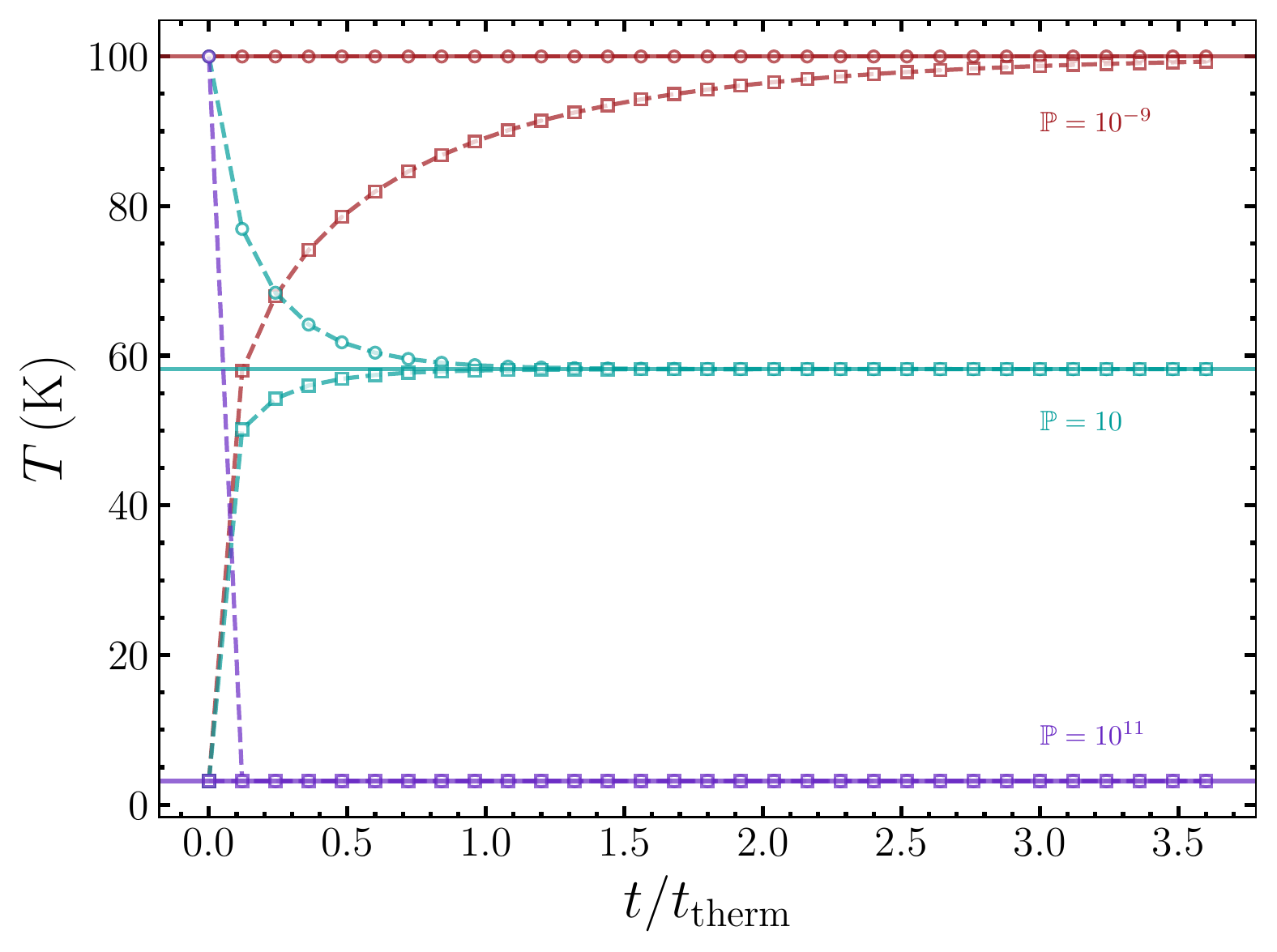}
    \caption{Evolution of the gas ($T_g$; circles) and radiation temperatures ($T_r$; squares) for the gas-radiation coupling test described in Section~\ref{sec:RadEqTest} as a function of time ($t$) scaled by the thermalisation timescale $t_\mathrm{therm} = 3.33 \times 10^{-11} \, \mathrm{s}$. We perform the test for values of  $\mathbb{P} = a_RT_{g,0}^4/(\rho RT_{g,0})$ of $10^{-9}$ (red), $10$ (cyan) and $10^{11}$ (violet), denoting varying levels of gas thermal inertia. We overplot, with solid lines, the final thermal equilibrium state  $T_{\mathrm{eq}}$ for each case, obtained from the solution to the fourth order polynomial Equation~\ref{eq:RadTempSteady}.}
    \label{fig:RadTempEq}
\end{figure}

In this problem, we test our treatment of the gas-radiation coupling term as implemented with a fixed-point Picard iteration scheme described in Section~\ref{sec:PicardUpdate}. We setup the problem in a fashion similar to \citet{Jiang_2021}, with an initial state where the gas and radiation temperatures are out of equilibrium ($|\ar T^4 - E_r| \geq 0$), to test whether the temperatures approach the correct state corresponding to thermal equilibrium ($a_RT^4 = E_r$). A uniform box is initialised in the region [0,1] cm, discretised with 512 cells, with a fixed specific opacity $\rho \kappa_P = 1  \, \mathrm{cm}^{-1}$ and an ideal gas $\gamma = 5/3$. The gas temperature is $T = T_{g,0} = 100 \, \mathrm{K}$ everywhere, and the radiation energy $E_{r,0} = 7.567 \times 10^{-13} \ergpcm$, which leads to a radiation temperature $T_{r,0} = (E_{r,0}/\ar)^{1/4} \approx 3.16 \, \mathrm{K}$. The boundary conditions are set to be zero-gradient outflow boundaries in $E_r$ and $F_r$. We test the setup with three different values of $\rho = 10^{-7}, 10^{-17}$, and $10^{-27} \gpcm$  respectively. We parameterise these three setups in terms of the dimensionless ratio $\mathbb{P} = a_RT_{g,0}^4/(\rho RT_{g,0})$ where $R$ is the ideal gas constant; our three cases correspond to values of $\mathbb{P}_0 = 10^{-9}, 10, $ and $10^{11}$ respectively. The three versions represent varying levels of thermal inertia of the gas, with a lower (higher) value of $\mathbb{P}_0$ indicating a higher (lower) gas thermal inertia, which means that the radiation (gas) temperature changes more significantly to reach the final equilibrium state. The tests are run up to a final time of $t = 5 t_{\mathrm{therm}}$, where $t_\mathrm{therm} = 1/(\rho \kappa_P c) = 3.33 \times 10^{-11} \, \mathrm{s}$ is the typical thermalisation timescale. We show the time evolution of $T$ and $T_r$ for the three cases in Figure~\ref{fig:RadTempEq}. We can compare the final thermal equilibrium state we obtain $T_{g,\mathrm{eq}} = T_{r,\mathrm{eq}} = T_{\mathrm{eq}}$ with that obtained analytically by enforcing total (gas+radiation) energy conservation in the initial and equilibrium states, i.e. 
\begin{equation}
    \label{eq:RadTempSteady}
    \frac{\rho R T_{g,0}}{\mu(\gamma -1)} + \ar(T_{r,0})^4 = \frac{\rho R T_{\mathrm{eq}}}{\mu m_{\mathrm{H}}(\gamma -1)} + \ar(T_{\mathrm{eq}})^4, 
\end{equation}
where $\mu = 0.6 m_{\mathrm{H}}$ is the mean particle mass of the gas, and $m_\mathrm{H}$ the mass of the hydrogen atom. The fourth order polynomial equation above can be solved to obtain $T_{\mathrm{eq}}$, and we also show this solution in Figure~\ref{fig:RadTempEq}. We find that the final state obtained in our numerical solution agrees with the analytically obtained value of $T_{\mathrm{eq}}$ to within $10^{-8}$ in all cases. In addition, we also verify that the total energy in our scheme is conserved to the precision of the implicit radiation update tolerance and/or the nonlinear Newton-Raphson update tolerance (whichever is higher). Overall, the results of this test demonstrate that the Picard iteration scheme described in Section~\ref{sec:PicardUpdate} captures the nonlinear gas-radiation coupling accurately.  

\subsection{Non-Equilibrium Marshak Wave}
\label{sec:MarshakWave}
The Marshak wave is a standard 1D non-equilibrium diffusion test originally proposed by \citet{Marshak_1958}, for which a semi-analytic solution has been provided by \citet{SuOlson_1996}. The domain is initially setup as a cold uniform medium ($T = 0$) with a constant absorption opacity $\sigma = \rho \kappa_P$, and at $t=0$ a constant radiation flux $F_r^{\mathrm{inc}}$ is applied at the $x=0$ boundary. The propagation of the radiation front heats the gas, and the time evolution is governed by the nonlinear equations of radiation diffusion and radiation-gas energy exchange. This is not a dynamical test, so the hydrodynamic evolution is switched off, with the exception of the thermal energy evolution due to radiation-gas energy exchange as described in Section~\ref{sec:PicardUpdate}. In addition, following \citet{SuOlson_1996}, we simplify the problem originally proposed by \citet{Marshak_1958}, in two ways. First, we adopt the Eddington approximation ($f_{xx} = 1/3$). Second, we adopt a specific heat capacity at constant volume for the fluid $C_v = \partial E_{\mathrm{int}}/\partial T = \alpha T^3$, where $E_{\mathrm{int}}$ is the gas internal energy and $\alpha$ is a fixed constant. The combination of these conditions allows a similarity transformation that converts the partial differential equations for the evolution of $E_r$ and $T$ into a system of ODEs, for which \citet{SuOlson_1996} provide a solution in terms of the dimensionless position $\chi = \sigma x$, and time $\tilde{t} = \epsilon c \sigma t$, where $\epsilon = 4a_R/\alpha$ is a fixed parameter. This solution is expressed in terms of the dimensionless radiation energy density $\mathcal{U}(\chi,\tau) = cE_r(x,t)/(4 F_r^{\mathrm{inc}})$ and gas temperature $\mathcal{V}(\chi, \tau) = ca_RT^4(x,t)/(4 F_r^{\mathrm{inc}})$. 

We simulate the problem on a one-dimensional grid of resolution $N_x = 1024$, on the domain $x \in [0,0.5] \, \mathrm{cm}$, with a uniform background density $\rho = 10^{-20} \gpcm$, and opacities $\kappa_P = \kappa_R = 4 \times 10^{21} \, \mathrm{cm}^2 \, \mathrm{g}^{-1}$, leading to a value of $\sigma = 40 \, \mathrm{cm}^{-1}$. The gas and radiation temperature are initialised to zero, and we use a value of $\epsilon = 4a_R/\alpha = 0.1$. The condition of a constant, half-isotropic incoming flux at the $x = 0$ boundary is imposed through the so-called Marshak boundary condition, given by the constraint 
\begin{equation}
    c E_r(0,t) + 2F(0,t) = 4 F_{\mathrm{inc}}
\end{equation}
where $E_r (0,t)$ and $F(0,t)$ are the values of the radiation energy and flux at the boundary wall/interface. We setup the boundary to mimic a source of radiation temperature $k_B T_\mathrm{inc} = 4.68 \times 10^{-13} \, \mathrm{eV}$, corresponding to $F_{\mathrm{inc}} = a_RcT_{\mathrm{inc}}^4/4 = 7.49 \times 10^{49} \, \mathrm{erg} \, \mathrm{cm}^{-2} \, \mathrm{s}^{-1}$. The other boundary ($x = 0.5$) is set to be reflective, though this choice does not matter since we halt the test before the advancing Marshak wave reaches it. We evolve the system at the light crossing timescale across a cell, up to a time corresponding to $\tilde{t} = 100$, corresponding to a physical time $t = 8.33 \times 10^{-10} \, \mathrm{s}$. 

\begin{figure*}
    \centering
    \includegraphics[width=0.98 \textwidth]{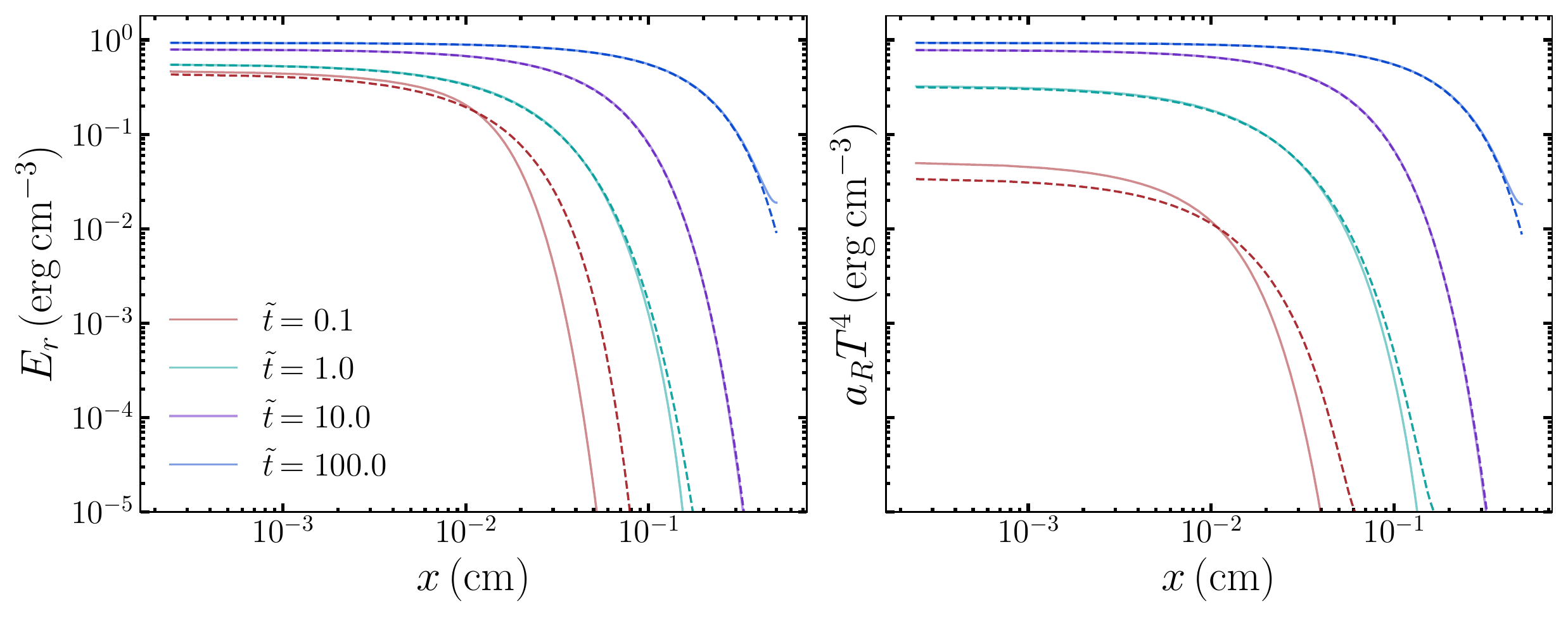}
    \caption{Results for the non-equilibrium Marshak wave test described in Section~\ref{sec:MarshakWave}. Solid lines indicate the simulation results for $E_r$ (left) and $a_RT^4$ (right) for dimensionless times $\tilde{t} = 0.1, 1, 10$ \& 100. Dashed lines indicate the corresponding solutions at these times obtained by numerically integrating the semi-analytic expressions of \citet{SuOlson_1996}.}
    \label{fig:Marshak_Wave}
\end{figure*}

We plot our simulation results for $E_r(x,t)$ and $a_RT(x,t)^4$ at $\tilde{t} = [0.1, 1, 10, 100]$ in Figure~\ref{fig:Marshak_Wave}. We compare this to the \citet{SuOlson_1996} solution for the dimensionless quantities  $\mathcal{U}(\chi,\tau)$ and $\mathcal{V}(\chi, \tau)$, which we compute using a publicly available code \footnote{\url{http://cococubed.asu.edu/research_pages/su_olson.shtml}} to numerically integrate their semi-analytic expressions with the parameters of our problem setup. We then rewrite these dimensionless quantities in terms of their dimensional counterparts. We find that our numerical solution reproduces the analytical quite accurately, especially at later times. At earlier times, the agreement is poor, as expected for a scheme that solves the full hyperbolic two-moment system of equations \citep[see,][for other two-moment schemes that report similar disagreements]{Gonzalez_2007,Skinner_2013,Tsang_2015}, since the \citet{SuOlson_1996} solution uses the diffusion approximation, which is inaccurate at early times when the wave has traversed an optical depth $\ll 1$; the error is that the diffusion approximation allows an infinite signal speed, while our two-moment scheme correctly captures the finite speed of light. This explains why our solution at early times lags the \citet{SuOlson_1996} solution; however, at these times our numerical solution is almost certainly more accurate. In any event, the very good agreement we obtain at late times, when \citeauthor{SuOlson_1996}'s diffusion approximation is accurate, shows that our scheme correctly reproduces the diffusion limit.

\subsection{Non-Equilibrium Radiation Shock}
The non-equilibrium radiation shock problem is a test of the non-equilibrium, fully coupled, radiation hydrodynamics system in the presence of shocks in an optically thick medium. This problem has been discussed in classical tests of radiation hydrodynamics \citep{Zeldovich_1967,Mihalas_1984}, with analytical solutions available under some physical conditions, namely cases where the material energy dominates the radiation energy, and/or under equilibrium conditions. More recently, however, \citet{Lowrie_2008} consider the non-equilibrium, high radiation energy density regime -- where radiation momentum/energy contributions are significant -- and provide a semi-analytic procedure to compute solutions for them. They show that the shock structure is fully determined by five parameters: i) the dimensionless pressure ratio $\mathbb{P}_0 = \ar T^4/(\rho a^2)$ where $T$, $\rho$ and $a$ are the temperature, gas density and adiabatic sound speed in the upstream regions, ii) the dimensionless specific opacity $\sigma_0 = \sigma_a L c/a$ where $\sigma_a \equiv \sigma_a(\rho,T)$ is the absorption specific opacity, and $L$ is the reference length scale iii) the dimensionless diffusion coefficient $D_0 = c/(3\sigma_t La)$ where $\sigma_t \equiv \sigma_t(\rho,T)$ is the total specific opacity, iv) the adiabatic index $\gamma$, and v) the upstream Mach number $\mathcal{M}_0 = v/a$ where $v$ is the velocity of the shock in the upstream region. 

In our test, we use the parameters $\gamma = 5/3$, a spatially uniform $\sigma_0 = 10^6$, $D_0 = 1$, $\mathbb{P}_0 = 10^{-4}$ and $\mathcal{M}_0 =3$ for the upstream state, from which we derive the corresponding physical conditions in cgs units, in which our code works. This set of parameters corresponds to a subcritical shock in which the pre-shock matter is preheated by the radiation to a temperature lower than the temperature in the downstream relaxation region. We initialise the problem in a domain $x \in [-0.0132,0.00255] \, \mathrm{cm}$, with the shock initially placed at $x = 0$. The upstream state of the gas ($x<0$) is set to be $\rho = 5.69 \, \gpcm$, $T = 2.18 \times 10^{6} \, \mathrm{K}$ and $v = 5.19 \times 10^7 \, \mathrm{cm} \, \mathrm{s}^{-1}$, and, by using the Rankine-Hugoniot jump conditions (by solving Equations 12 and 13 of \citet{Lowrie_2007}), we obtain the downstream state $\rho = 17.1 \, \gpcm$, $T = 7.98 \times 10^{6} \, \mathrm{K}$ and $v = 1.73 \times 10^7 \, \mathrm{cm} \, \mathrm{s}^{-1}$. The absorption specific opacity is fixed at $\rho \kappa_P = \rho \kappa_R = 577 \, \mathrm{cm}^{-1}$ everywhere in the domain, enforced by setting $\kappa_P = \kappa_R = 577/\rho \, \mathrm{cm}^{2}\, \mathrm{g}^{-1}$. This condition on the opacity, although unphysical, is enforced to mimic the solutions provided in \citet{Lowrie_2008}. The gas and radiation are initialised to be in equilibrium at $t=0$, and we use an ideal monoatomic gas EOS ($\gamma = 5/3$) with a mean particle mass $\mu = m_{\mathrm{H}}$. We evolve the system to a time $t = 10^{-9} \, \mathrm{s}$, corresponding to about 3 crossing times of the computational domain. The boundary conditions at the lower (higher) $x$ boundary is fixed to the asymptotic downstream (upstream) state of the shock provided in the initial conditions. We also use the Eddington approximation ($f_{\mathrm{xx}} = 1/3$) for this problem to allow comparison with the semi-analytical solution derived under the same assumption. The grid is discretised with a base grid resolution of 640 cells, and adaptively refined on the gas temperature, using the default refinement condition in \verb|FLASH| based on a modified second derivative of a variable \citep{Fryxell_2000}, to a maximum refinement level of $l_{\mathrm{max}} = 3$, corresponding to a maximum resolution of 2560 cells. We use the modified CFL timestep criterion (Equation~\ref{eq:timestep}) for this problem, with a CFL number $C_0 = 0.5$. 

\begin{figure}
    \centering
    \includegraphics[width=\columnwidth]{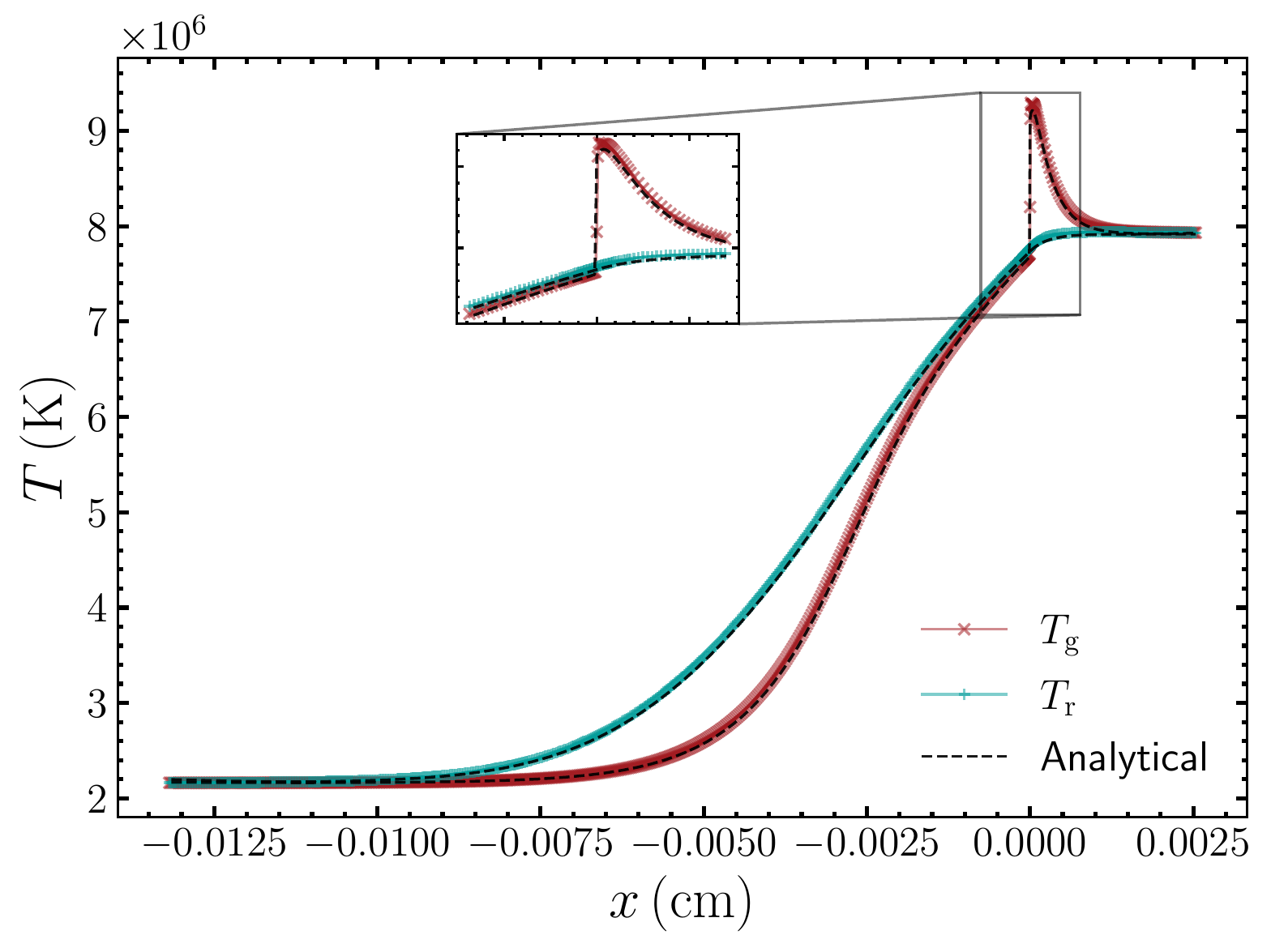}
    \caption{Gas ($T_{\mathrm{g}}$; red crosses) and radiation ($T_r$; green plus signs) temperatures for a subcritical non-equilibrium radiative shock; the inset shows the Zel'Dovich spike in detail. The semi-analytical solution of \citet{Lowrie_2008} is overplotted with black dashed lines.}
    \label{fig:Radiation_Shock}
\end{figure}

In Figure~\ref{fig:Radiation_Shock} we plot the numerical solution we obtain for the gas ($T_{\mathrm{g}}$) and radiation ($T_r$) temperatures, with the inset showing the so-called Zel'Dovich spike -- an inherently non-equilibrium feature -- in further detail. We overplot the solution obtained with the \citet{Lowrie_2008} semi-analytical procedure as well for comparison. We find very good agreement between the two solutions, with a relative error in the $L_1$ norm of $\sim 0.6\%$ in $T$ and $T_r$, and find that the sharp temperature spike is well-captured by our refined domain. This test demonstrates that our scheme is able to accurately capture fully coupled radiation-gas dynamics in the presence of strong discontinuities. 

\subsection{Advecting Radiation Pulse}
To test the accuracy of our implementation of the $\mathcal{O}(\beta \tau)$ relativistic correction source terms that arise from the mixed frame formulation of the RHD moment equations, in a fully-coupled RHD problem, we simulate the test described by \citet{Krumholz_2007a}. The test involves the advection of a pulse of radiation energy in an optically thick gas, with a uniform background flow velocity. The initial condition is such that the system is in both pressure and radiative equilibrium everywhere, but with a Gaussian-shaped pulse centred at $x=0$ within which there is a local increase in the temperature and radiation pressure, and a corresponding decrease in the gas pressure and density. At times $t>0$, radiation diffuses out of the pulse, leading to the loss of pressure balance, and the gas starts to move into the region occupied by the pulse. While a time-dependent solution is not known analytically, the problem is nonetheless a useful test if we perform two cases of the setup: one where the gas is initially at rest ($v=0$), and another where the gas is provided an initial uniform velocity ($v=v_0$). If the velocity dependent terms are included correctly, the solutions for the two cases should be identical to each other except for displacement by a distance $v_0 t$. 

To setup this problem, we initialise the temperature as a function of position to
\begin{equation}
    \frac{T}{T_{0}}=1+\exp \left(-\frac{x^{2}}{2 w^{2}}\right) ,
\end{equation}
where $w=24$ cm is the pulse width and $T_0 = 10^7 \, \mathrm{K}$ is the background temperature. Imposing the conditions of radiative and pressure equilibrium everywhere immediately gives the corresponding gas density, 
\begin{equation}
    \rho=\rho_{0} \frac{T_{0}}{T}+\frac{a_{\mathrm{R}} \mu}{3 k_{\mathrm{B}}}\left(\frac{T_{0}^{4}}{T}-T^{3}\right),
\end{equation}
where $\rho_0 = 1.2 \, \gpcm$ is the background gas density, and $\mu = 2.33 m_{\mathrm{H}}$ is the mean particle mass. We use a spatially uniform gray opacity of $\kappa_P = \kappa_R = \kappa_0 = 100 \, \mathrm{cm}^2 \, \mathrm{g}^{-1}$, and use a value of $v_0 = 10 \, \mathrm{km} \, \mathrm{s}^{-1}$ for the moving pulse case. The simulation domain goes from $-512$ to 512 cm and is resolved by 1024 uniformly spaced cells. The Eddington tensor is assumed to be spatially and temporally uniform with a value $f_{xx} = 1/3$. Periodic boundary conditions are used on the radiation and gas, and the system is evolved to a final time of $t = 2w/v_0 = 4.8 \times 10^{-5} \, \mathrm{s}$, so the pulse is advected by twice its initial width. We use the modified CFL timestep criterion (Equation~\ref{eq:timestep}) for this problem, with a CFL number $C_0 = 0.4$. 

\begin{figure}
    \centering
    \includegraphics[width=\columnwidth]{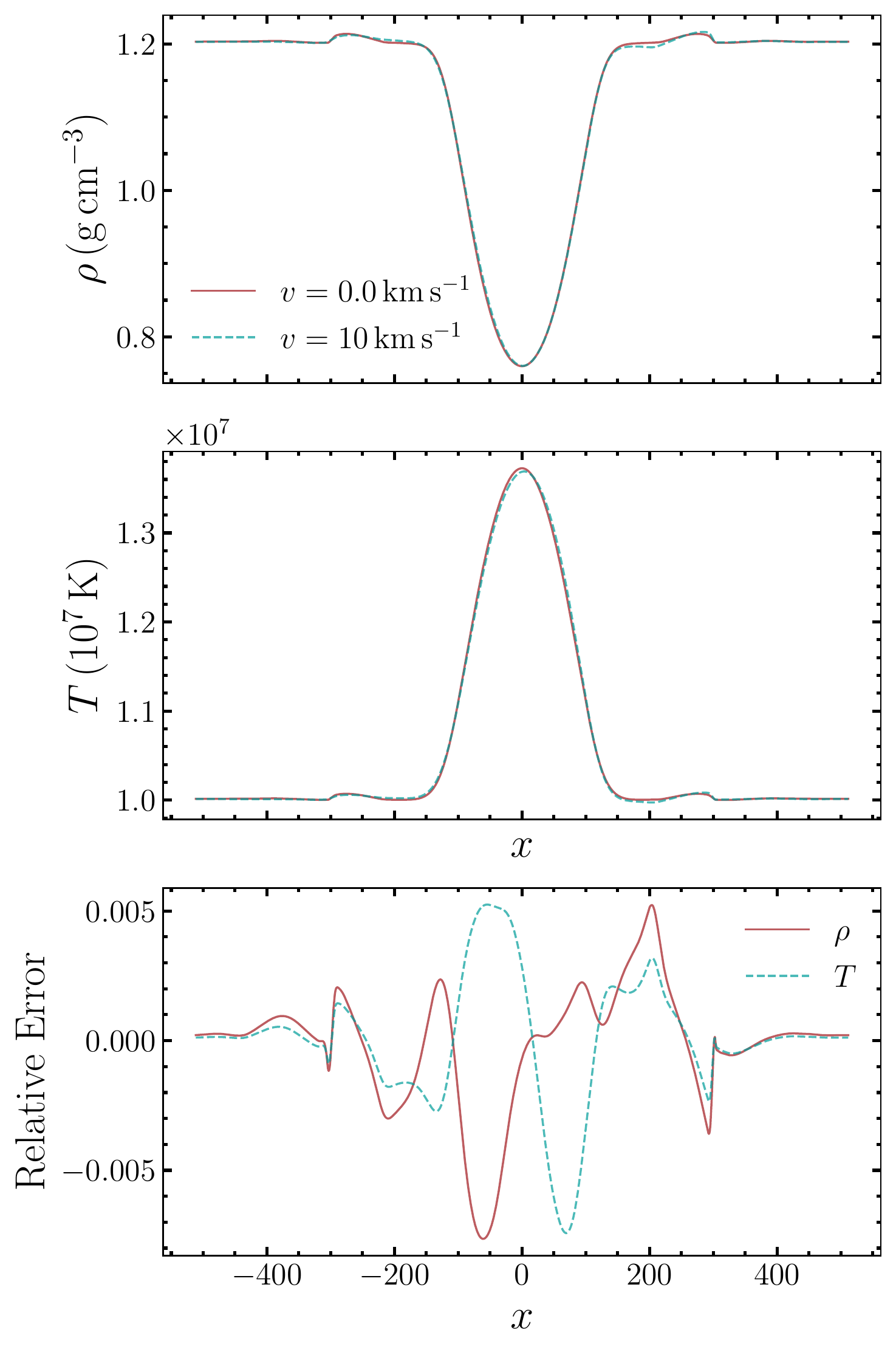}
    \caption{Plot comparing the density $\rho$ (top) and the gas/radiation temperature $T$ (middle panel) profiles we obtain with our scheme for an initially unadvected (red) and advected (cyan) gas with the radiating pulse advection test of \citet{Krumholz_2007a}. The bottom panel shows the relative error between the two cases for $\rho$ (red) and $T$ (cyan), which we find to be bounded by 0.7\% over the domain.  }
    \label{fig:Advecting_Test}
\end{figure}

In Figure~\ref{fig:Advecting_Test}, we compare the results of the two runs; for the advected case we have shifted the solution by a distance $v_0 t = 48$ cm in the $-x$ direction, so that it should lie on top of the unadvected case. We see that the agreement between the advected and unadvected solutions is very good. The maximum relative errors are bounded by 0.7 \% over the domain, and we obtain a relative $L_1$ norm error of 0.1\%. This demonstrates that our scheme is handling the advection of radiation by gas in the diffusion regime appropriately, and provides evidence for the correct modelling of the velocity-dependent radiative work and advection terms in the moment equations. 

\subsection{Spherical Expansion of Radiation-Pressure Dominated $\ion{H}{ii}$ region} 
\label{sec:expandingshell}
In our next test, we simulate the radiation pressure-driven expansion of a thin, dusty, spherical shell as given in \citet{Skinner_2013} (\citetalias{Skinner_2013} hereafter). The problem considers an idealised central source of photons -- for example a massive star or star cluster -- impinging on a surrounding dusty cloud that absorbs the photons, transferring momentum to the gas. This is a fully-coupled radiation-hydrodynamic problem that tests gas-radiation momentum exchange in three dimensions. The central source function for the radiation field is given by 
\begin{equation}
    \label{eq:jstar}
    j_{*}(r)=\frac{L_{*}}{\left(2 \pi R_{*}^{2}\right)^{3 / 2}} \exp \left(-\frac{r^{2}}{2 R_{*}^{2}}\right), 
\end{equation}
where $L_*$ is the luminosity of the cluster and $R_*$ the size of the source. We add $j_*$ as a source term on the right hand side of our equation for $E_r$. In addition, we add the corresponding term $j_*(r)/(4 \pi \rho \kappa_P)$ to the source function of the ray-tracer, taking into account the contribution of the central source in the computation of the Eddington tensor $\mathbfss{T}$. The test assumes that at $t=0$, a shell of thickness $H$ and zero velocity is present at a radius $r = r_0$, and monitors the evolution of the shell radius and velocity with time. The radial density profile at $t=0$ is given by 
\begin{equation}
    \rho_{\mathrm{sh}}(r)=\frac{M_{\mathrm{sh}}}{4 \pi r^{2} \sqrt{2 \pi R_{\mathrm{sh}}^{2}}} \exp \left(-\frac{\left(r-r_{0}\right)^{2}}{2 R_{\mathrm{sh}}^{2}}\right), 
\end{equation}
where $M_{\mathrm{sh}}$ is the gas mass in the thin shell, and $R_{\mathrm{sh}} \equiv H/(2\sqrt{2 \ln 2})$ is the half-width of the shell. The dust opacity $\kappa_0$ is set to be constant in space and time for simplicity. Following \citetalias{Skinner_2013}, we define the following quantities to non-dimensionalise the problem setup: a length unit of $r_0$, density unit $\rho_0 = 3M_{\mathrm{sh}}/(4\pi r_0^3)$, velocity unit $a_0$ corresponding to the isothermal sound speed, and time unit $t_0 = r_0/a_0$. Under the thin-shell approximation, and assuming reprocessed radiation pressure in the dusty shell to be the only source of radial pressure, it is possible to express the time evolution of the radius of the shell as an analytic parametric equation in these dimensionless units (Equation 106 of \citetalias{Skinner_2013}), given by 
\begin{equation}
    \label{eq:ShellR_An}
    \tilde{t}=\frac{1}{\mathcal{M}_{0} \sqrt{2}}[\sqrt{\tilde{r}} \sqrt{\tilde{r}-1}+\ln (\sqrt{\tilde{r}}+\sqrt{\tilde{r}-1})],
\end{equation}
where $\tilde{t} = t/t_0$, $\tilde{r} = r/r_0$ and $\mathcal{M}_{0} =  \sqrt{ L_* \kappa_0/(4 \pi r_0ca_0^2)}$ is the reference dynamical Mach number. Similarly the shell velocity is given by (Equation 105 of \citetalias{Skinner_2013})
\begin{equation}
    \label{eq:ShellVel_An}
    \frac{d \tilde{r}}{d \tilde{t}}=\mathcal{M}_{0} \sqrt{2}\left(1-\frac{1}{\tilde{r}}\right)^{1 / 2}. 
\end{equation}
We use these relations to compare with the shell radius and velocity in our simulations below. \citetalias{Skinner_2013} simulate the problem with an isothermal equation of state, and under the conditions of radiative equilibrium ($|a_RT^4-E_r| = 0$) for simplicity, which we also adopt here. 

We setup the problem with parameters identical to those specified in \citetalias{Skinner_2013}: initial shell radius $r_0 = 5\, \mathrm{pc}$, thickness $H = 0.3 r_0 = 1.5 \, \mathrm{pc}$, central source luminosity $L_* = 1.989 \times 10^{42} \, \mathrm{erg} \, \mathrm{s}^{-1}$, central source size $R_* = r_0/8 = 0.625 \, \mathrm{pc}$, and dust opacity of $\kappa_P = \kappa_R = \kappa_0 = 20 \, \mathrm{cm}^{2} \, \mathrm{g}^{-1}$. The isothermal sound speed is set to $a_0 = 2 \, \mathrm{km} \, \mathrm{s}^{-1}$, which corresponds to a gas temperature $T\sim 481 \, \mathrm{K}$ assuming a mean particle mass $\mu = m_\mathrm{H}$. The simulation is performed on the domain $(x,y,z) \in [-10,10]^3 \, \mathrm{pc}$, with the source at $x=y=z=0$, with outflow boundary conditions on the gas and radiation. We note that \citetalias{Skinner_2013} simulate only a quadrant of the sphere with reflecting boundary conditions at the $x=0$ boundary, which we did not repeat here to avoid having to implement reflecting boundary conditions in the RT solver used to compute $\mathbfss{T}$. We use AMR for this test, with a base resolution of $32^3$, and allow up to four levels of refinement, corresponding to an effective resolution of $256^3$. We refine blocks where $a_0/\sqrt{G \rho} > 16 \Delta x$. We note that this is identical to the standard Jeans refinement criteria, but we remind the reader that we do not have self-gravity in this simulation. We also perform simulations on uniform grids of resolution $64^3$, $128^3$, and $256^3$\footnote{\citetalias{Skinner_2013} perform their simulations at this resolution} to study the dependence of shell evolution on resolution.  

\begin{figure}
    \centering
    \includegraphics[width=\columnwidth]{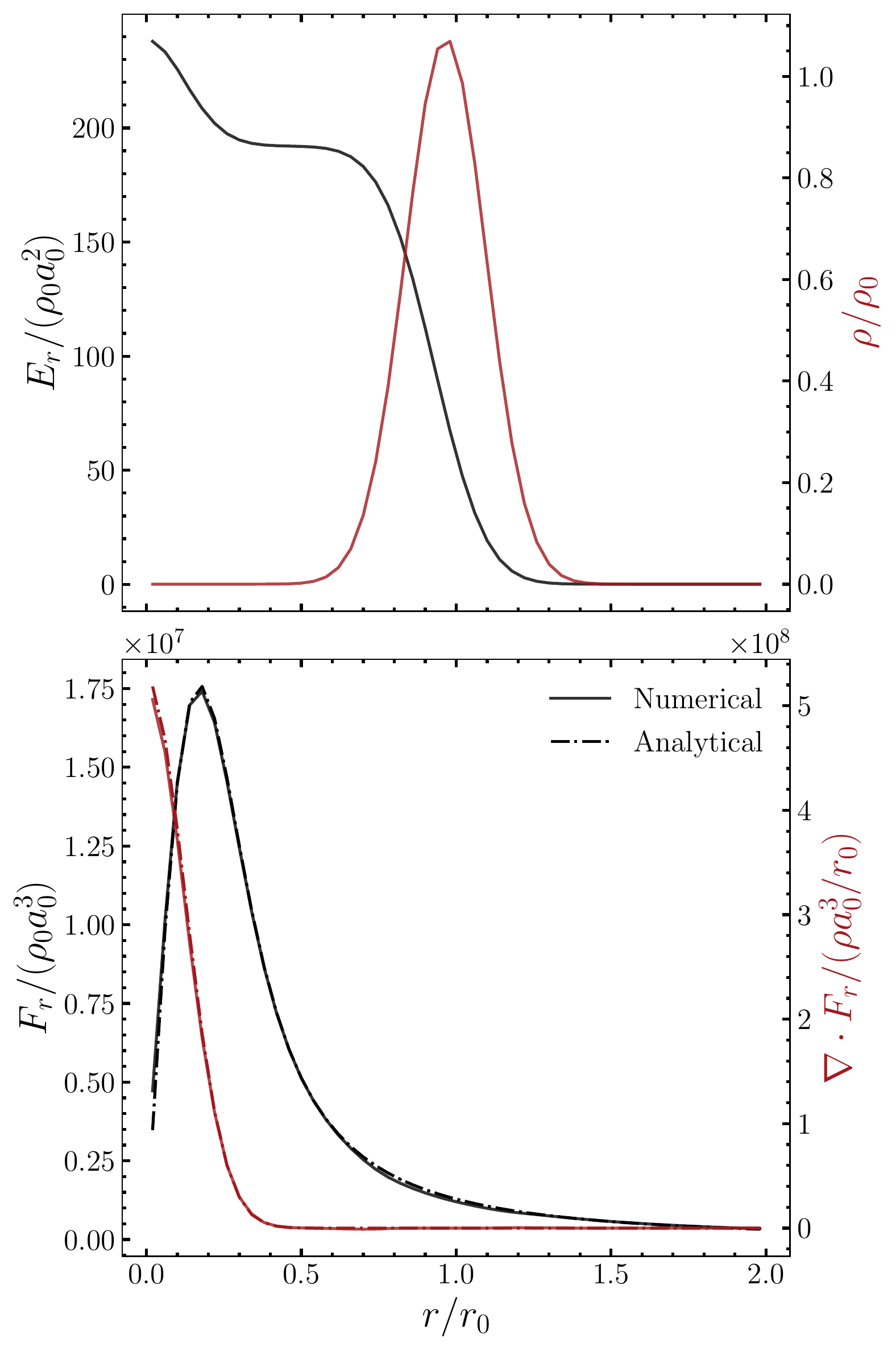}
    \caption{Steady state radial profiles of the radiation energy density $E_r$ (top panel; black), gas mass density $\rho$ (top panel; red), radiation flux $F_r$ (bottom panel; black), and the divergence of the radiation flux $\nabla \cdot F_r$ (bottom panel; red) for the radiation-driven thin shell test without hydrodynamical evolution. The quantities are expressed in the dimensionless units of the problem described in the text. We overplot the analytically derived profiles of $\nabla \cdot F_r$ and $F_r$ given in Equations~\ref{eq:DivFstarThinshell} and ~\ref{eq:FstarThinShell} in the bottom panel with dot-dashed lines, and verify that our solutions are in agreement with them. These profiles represent the initial conditions for the subsequent dynamical evolution of the shell (see text for details).}
    \label{fig:thinshell_radprofiles}
\end{figure}

While we initialise the density distribution following \citetalias{Skinner_2013}, we must use a different method to initialise the radiation energy density and flux, due to the difference in closures between \texttt{VETTAM} and \citetalias{Skinner_2013}'s $M_1$ approach. \citetalias{Skinner_2013} initialise the problem with a quasi-static steady state radiation energy density ($E_r^* (r)$) and flux ($F_r^* (r)$), derived under the condition of radiative equilibrium. Specifically, they estimate $F_r^* (r)$ by setting $\partial E_r/\partial t = 0$, which gives
\begin{equation}
    \label{eq:DivFstarThinshell}
    \nabla \cdot F_r^* = j_*(r), 
\end{equation}
which can be inverted to obtain 
\begin{equation}
    \label{eq:FstarThinShell}
    F_{*}(r)=\frac{L_{*}}{4 \pi r^{2}}\left[\operatorname{erf}\left(\frac{r}{\sqrt{2} R{*}}\right)-\frac{2 r}{\sqrt{2 \pi R{*}^{2}}} \exp \left(-\frac{r^{2}}{2 R_{*}^{2}}\right)\right].
\end{equation}
Similarly, they obtain the solution for $E_r^* (r)$ by setting $\partial F_r/\partial t = 0$, which gives $\nabla \cdot \mathbfss{P} = -\rho \kappa_0 \mathbfit{F}/c$, and then invoking the $M_1$ closure to relate $\mathbfss{P}$ to $E_r$. Since we do not have an analytic closure relation, our alternative approach is to set $E_r = F_r = 0$ as the initial condition, and evolve the system without hydrodynamics for a transient period until the radial profiles of $E_r$ and $F_r$ reach a steady state. We plot the radial profiles of $E_r$, $F_r$ and $\nabla \cdot F_r$ obtained at the steady state in our simulation in Figure~\ref{fig:thinshell_radprofiles}; we also show the profile of density $\rho$ for reference. We find that $\nabla \cdot F_r$ and $F_r$ are very close to the results given by Equations~\ref{eq:DivFstarThinshell} and \ref{eq:FstarThinShell} respectively, indicating that the $M_1$ approximation is close to our full VET result for this problem. We also verified that the solution converges to this steady state solution from other initial conditions as well.  

\begin{figure*}
    \centering
    \includegraphics[width=0.98 \textwidth]{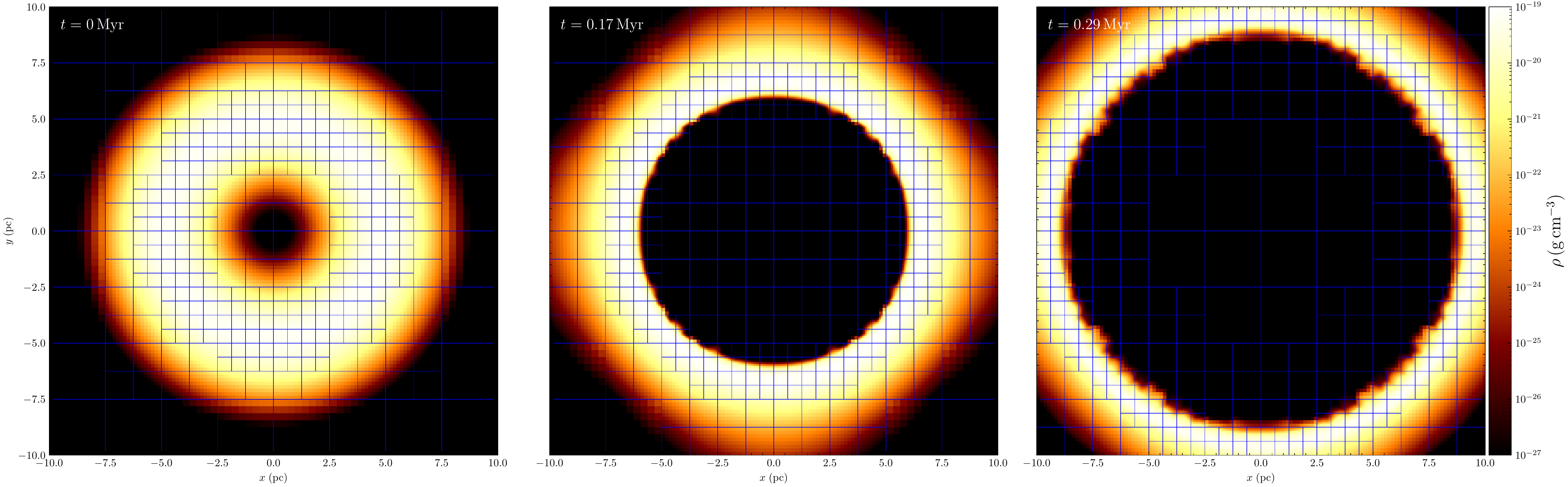}
    \caption{Slice plots of the gas density $\rho$ at times $t =0$, $0.17$ and $0.29 \, \mathrm{Myr}$ for the radiation-driven shell evolution test simulation, with the block structure of the AMR domain overplotted.}
    \label{fig:thinshell_slices}
\end{figure*}

\begin{figure}
    \centering
    \includegraphics[width=\columnwidth]{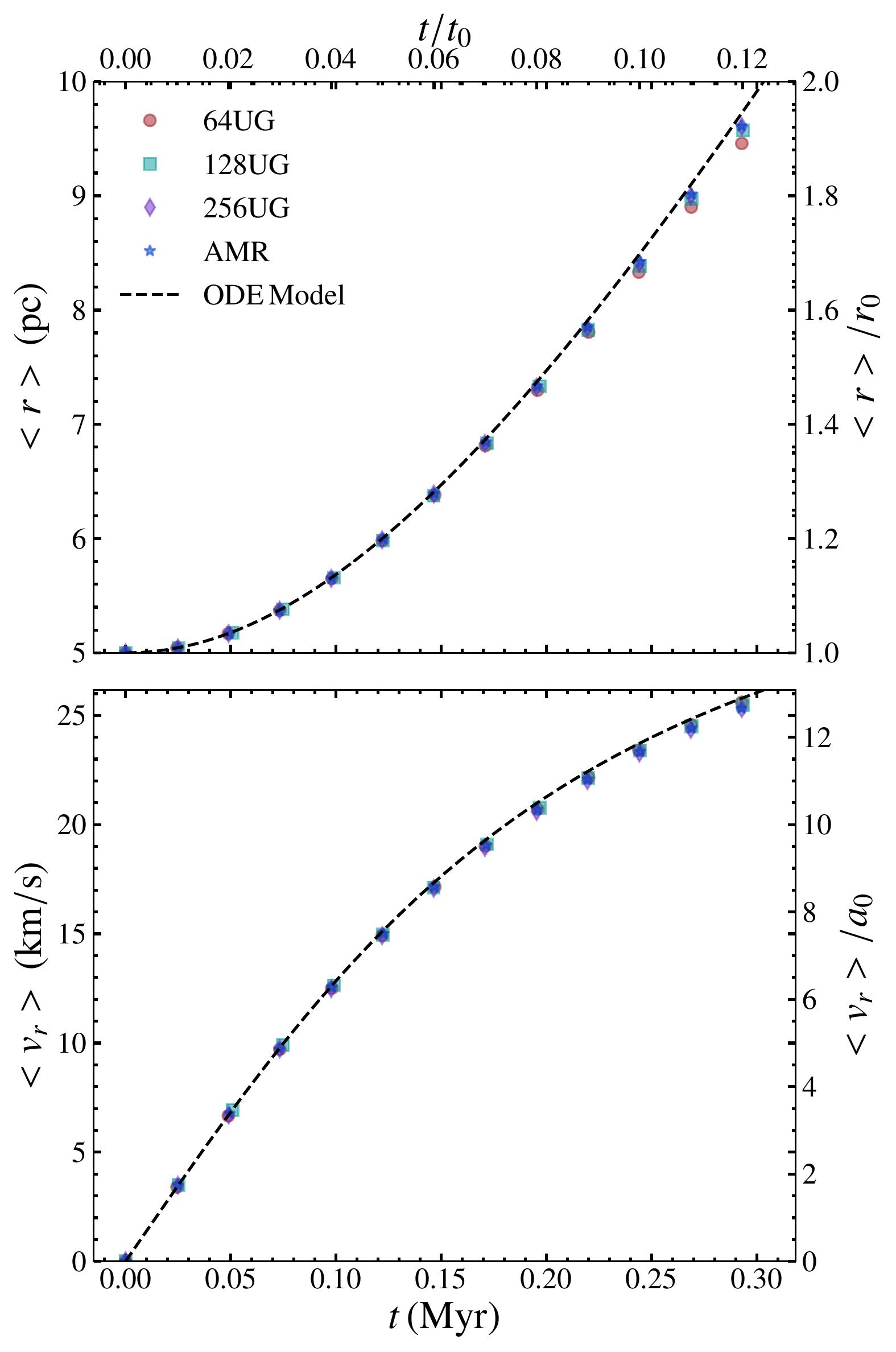}
    \caption{Evolution of the mass-weighted average radius $\langle r\rangle$ (Equation~\ref{eq:mwradius}; top) and radial velocity $\left\langle v_{r}\right\rangle$ (Equation~\ref{eq:mwvel}; bottom) obtained for the radiation-driven thin shell expansion test at regular time intervals, and compared at different resolutions. In addition, we plot the analytical solutions for $\langle r\rangle$ and $\left\langle v_{r}\right\rangle$ from the ODE model of \citet{Skinner_2013} for comparison.}
    \label{fig:thinshell_evolution}
\end{figure}

Once the radiation field has reached steady state, we turn hydrodynamics back on, and allow the system to evolve. We use the unmodified CFL condition to determine the timestep, enforcing a density floor of $\rho_{\mathrm{min}} = 10^{-8} \rho_0$ to prevent very small timesteps, and run the simulation to a final time of $t_{\mathrm{final}} = 0.5 \, \mathrm{Myr}$. We show slice plots following the evolution of the expanding thin shell for three different times in Figure~\ref{fig:thinshell_slices}, with the AMR block structure overplotted. We then estimate the radius of the shell at a given time $t$ by calculating the mass-weighted average radius in our computational domain, given by 
\begin{equation}
    \label{eq:mwradius}
    \langle r\rangle \equiv \frac{\int \rho r d V}{\int \rho d V},
\end{equation}
where $r = \sqrt{x^2+y^2+z^2}$ is the radius of a grid point in the domain, $\rho$ the local density, and $dV$ the volume of the cell. In addition, we can also compute the mass-weighted radial velocity as 
\begin{equation}
    \label{eq:mwvel}
    \left\langle v_{r}\right\rangle \equiv \frac{\int \rho(\mathbfit{v} \cdot \hat{\mathbfit{r}}) d V}{\int \rho d V}, 
\end{equation}
where $\mathbfit{v} \cdot \hat{\mathbfit{r}}$ denotes the Cartesian velocities projected in the radial direction. We show the time evolution of our computed values of $\langle r\rangle$ and $\left\langle v_{r}\right\rangle$ in Figure~\ref{fig:thinshell_evolution} for our fiducial AMR simulation, and the uniform grid versions at different resolutions. These are compared with the analytical relations for the shell radius and velocity evolution given by Equations~\ref{eq:ShellR_An} and ~\ref{eq:ShellVel_An} respectively. We find excellent agreement at all resolutions, although, as expected, better agreement at higher resolutions. The maximum error in the solutions for the radius (velocity) are bounded by 3.7\% (5\%), 2.1\% (2.2\%) and 1.4\% (1.9\%) for the uniform grid $64^3$, $128^3$ and $256^3$ versions respectively. The maximum errors in the $256^3$ effective resolution AMR version are 1.3\% and 1.8\% for the radius and velocity respectively, which is comparable to the errors obtained by the $256^3$ uniform run. However, we find that the AMR run uses about 30\% less CPU time than the uniform grid run, and is thus more efficient.

\subsection{Shadow Tests}
Shadow tests in various forms are commonly used to investigate how well RHD schemes reproduce and preserve angular variations in the radiation field in the presence of a mixture of optically thin and thick gas, and to illustrate the varying levels of directional accuracy that different closure methods achieve. For instance, it is well known that methods based on the diffusion approximation ($\mathbfss{P} = (1/3) E_r \mathbfss{I}$) fail to cast shadows. Local closures such as the $M_1$ approximation fail to propagate the radiation field correctly in the presence of multiple sources of radiation (or effectively, convergent rays) in an optically thin medium \citep[as demonstrated in tests by][]{Rosdahl_2013,Kannan_2019}. We perform three forms of shadow test below to demonstrate that our nonlocal VET-closed scheme can cast shadows correctly, even in situations where other methods fail. Hydrodynamic evolution is switched off in all three shadow tests (i.e. $\mathbf{v} = 0$).

In our first version, we perform a test similar to that first presented in \citet{Rijkhorst_2006}, and recently shown by \citet{Klassen_2014} and \citet{Rosen_2017}, to demonstrate the shadow cast by an optically thick cloud in an optically thin medium when irradiated by two point sources of radiation, where the point source contributions are handled by a ray-tracer, and the subsequent diffuse re-emission with the moment method. We show this as a demonstration of the workings of our hybrid radiation algorithm described in Section~\ref{sec:sinkcontribution}. While this is a useful problem for testing the coupling between the point and diffuse sources of radiations, the setup of the test is such that the direct irradiation on the clump is the agent that casts the shadow, whereas the moment method is only used for the diffuse re-emission that is largely isotropic. In other words, the presence of the shadow in this case is largely insensitive to the closure for the moment method adopted. 

With this in mind, in the next two versions, we instead model setups with solely diffuse sources of radiation that are handled by the moment method only. This is important to test, because in a dynamical simulation, there can self-consistently arise sources of radiation that cannot be reduced to a point source (or sink particle) -- for instance, heated overdensities in a clumpy, dusty medium -- and whose contribution to the energy budget of the gas could be significant. While the geometric distributions of diffuse sources could be quite general in a dynamical simulation, we consider only simple cases here. For our first test with diffuse sources, we use a modified version of the hybrid radiation test setup, but with the point sources replaced by diffuse spherical sources of radiation modelled by a Gaussian source term. This should, qualitatively, cast a shadow similar to the hybrid radiation test, if the moment method used can handle the propagation of radiation in such a setup correctly. We show that our method passes this test in Section~\ref{sec:diffuseshadow}. Following that, in Section~\ref{sec:diffuseextended}, we present a test where we replace the point sources with an extended, non-spherical source of radiation that might be representative, for example, of emission from a hot, clumpy, filament in an otherwise optically thin medium, or from a geometrically thin accretion disk in an optically thin atmosphere. We again show that a qualitatively correct shadow is obtained with our scheme for this setup. We elaborate on the test setups, and show the results we obtain, below.  

\subsubsection{Hybrid Radiation with Point Sources}
\label{sec:hybridshadow}
First, we perform a test with the hybrid radiation algorithm to demonstrate that the coupling between the direct radiation field modelled with the ray-tracer, and the reprocessed radiation that we handle using the VET closure is implemented correctly. To setup this test, we place a dense clump of material at the centre of a $(2000 \, \mathrm{AU})^3$ computational domain, with radius $267 \, \mathrm{AU}$ and density $\rho_\mathrm{c} = 3.89 \times 10^{-17} \gpcm$. The clump is surrounded by an optically thin ambient medium with density $\rho_\mathrm{a} = 3.89 \times 10^{-20} \gpcm$. The gas temperature is taken to be spatially uniform with a value of 20 K. The clump is irradiated by two point sources of solar luminosity (i.e. $1 \, L_{\sun}$) placed 368 AU from the edges of the clump at $(x,y,z) = (\pm 635,-635,0) \, \mathrm{AU}$. The opacity for the direct stellar radiation is set to $\kappa_* = 64 \, \mathrm{cm}^2 \, \mathrm{g}^{-1}$, whereas the moment method uses the gray opacities from \citet{Semenov_2003}. We use AMR for this test, discretising the grid with a base resolution of $128^3$, and refine the grid based on the modified second derivative condition in FLASH \citep{Fryxell_2000} on the variables $E_r$ and $\rho$, up to a maximum resolution of $512^3$. In addition, we ensure that the dense clump is always refined to the maximum resolution. We use 48 angles for the ray-tracer, although we found that our results were insensitive to this choice.

\begin{figure*}
    \centering
    \includegraphics[width=\textwidth]{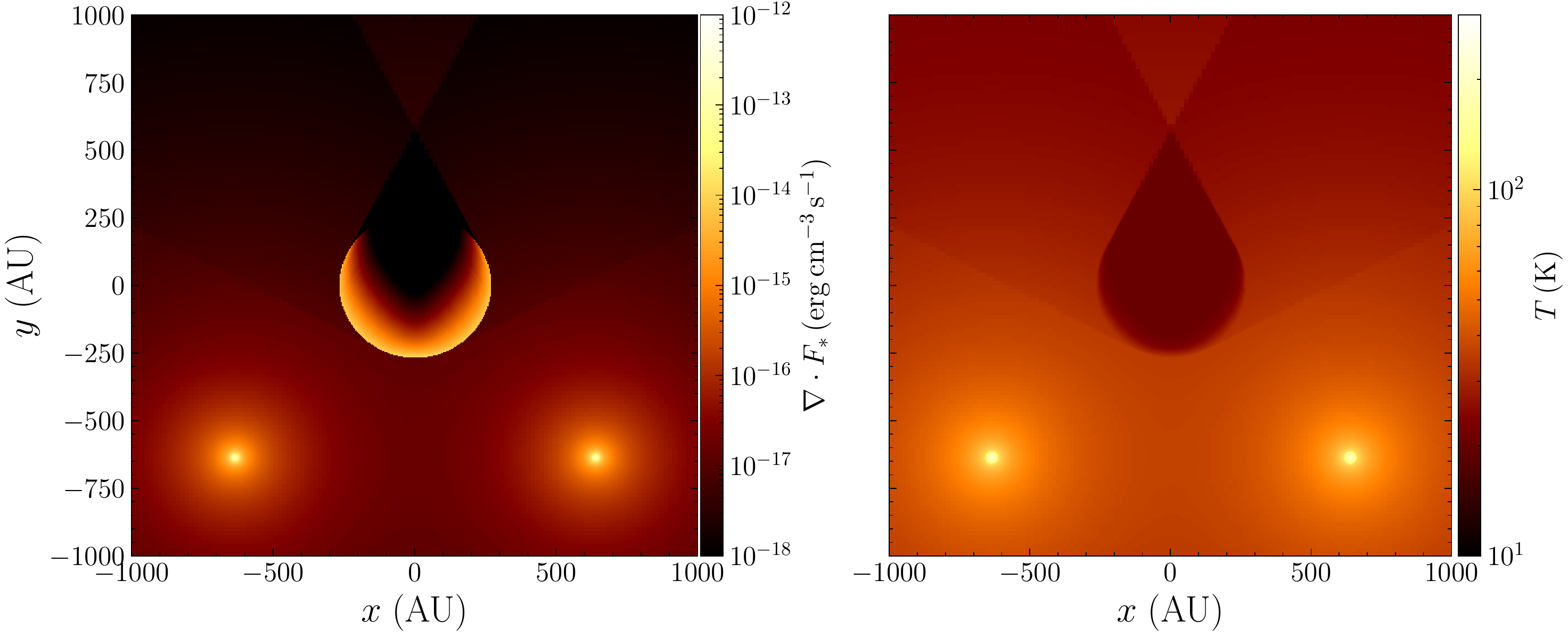}
    \caption{Slice plot of the heating rate per unit volume $\nabla \cdot F_*$ (left) from two point sources of luminosity $L_* = 1 \, L_{\sun}$ irradiating an optically thick clump, and the resulting gas temperature $T$ (right) obtained through absorption and subsequent re-emission of this energy. This test demonstrates the working of our hybrid radiative transfer scheme (see Section~\ref{sec:hybridshadow}) .}
    \label{fig:hybridshadow}
\end{figure*}

In Figure~\ref{fig:hybridshadow} we show the irradiation from the direct field from point sources (left), and the subsequent gas temperature obtained after this is reprocessed by the gas and it cools. As we can see, our point source irradiation produces a clear shadow, whereas the diffuse re-emission works to smooth the temperature field. This test shows that our hybrid approach of coupling the contribution of radiation from point sources and the diffuse emission works correctly. 

\subsubsection{Spherical Diffuse Sources}
\label{sec:diffuseshadow}
\begin{figure}
    \centering
    \includegraphics[width=\columnwidth]{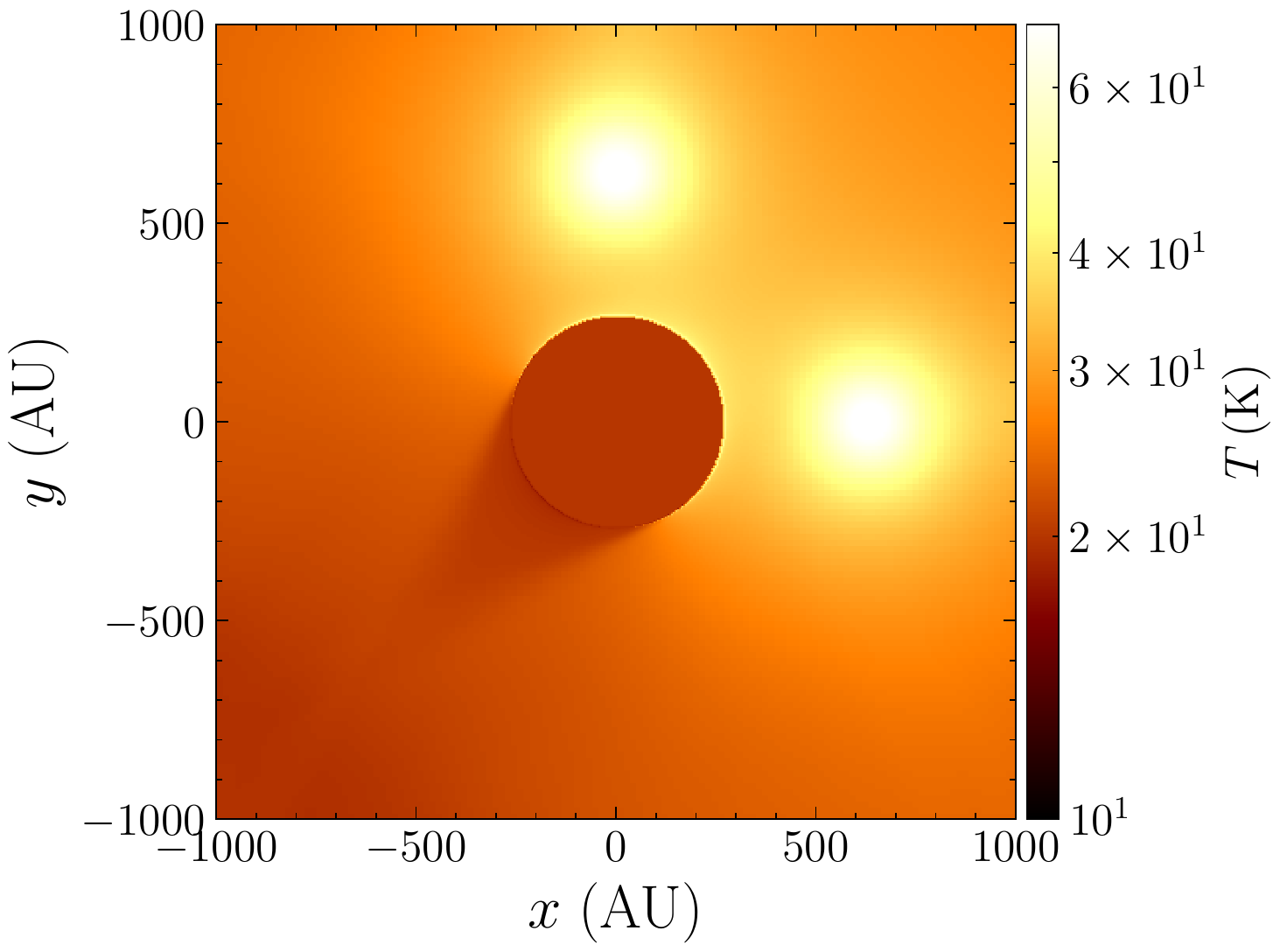}
    \caption{Slice plot of the gas temperature $T$ at a time corresponding to one crossing time of the computational box, demonstrating the shadow cast by two diffuse spherical sources of radiation impinging on an opaque clump of gas, lying in an optically thin ambient medium  (see Section~\ref{sec:diffuseshadow}) .}
    \label{fig:diffuseshadow}
\end{figure}

To demonstrate that our VET method is capable of capturing shadows even in the presence of purely diffuse sources of radiation, we setup a modified version of the test described in the previous section. The primary modification we make is to replace the point sources of radiation with diffuse sources, modelled with a Gaussian source function $j_*(r)$ with a profile identical to that given in Equation~\ref{eq:jstar}. We add this as a source term for our equation of $E_r$ in our VET scheme. We use a value of $L_* =  10 \, L_{\sun}$ and $R_* = 54 \, \mathrm{AU}$ for both the sources. We also change the positions of the sources with respect to the previous test such that they are $90^{\deg}$ apart with respect to each other, at $(635,0,0) \, \mathrm{AU}$ and $(0,635,0) \, \mathrm{AU}$ respectively. We make this change simply to prevent confusion with the test described in the previous version. In addition, for simplicity, we set the diffuse radiation opacities to be independent of the gas state with a constant value of $\kappa_P = \kappa_R = 100 \, \mathrm{cm}^{2} \, \mathrm{g}^{-1}$. This constant value ensures that the clump is optically thick to the diffuse radiation, whereas the ambient medium is optically thin to it. We use a base grid resolution of $128^3$ for this test, and refine based on the modified second derivative condition in \verb|FLASH| \citep{Fryxell_2000} on the variables $\rho$ and $j_*$, up to a maximum resolution of $512^3$, which ensures that the sources and the edges of the clump are well resolved. In addition, we use a total of 192 angles for the ray-tracer while calculating the VET. In Figure~\ref{fig:diffuseshadow} we show a slice plot of the temperature structure obtained with our scheme at a time corresponding to a light crossing time of the box. We can see that a clear shadow is cast by the optically thick clump. We point out that this is a challenging test due to the presence of converging rays of radiation, and local closure methods would fail to propagate the radiation correctly in such a setup, as we shall demonstrate in Section~\ref{sec:compareschemes}.

\subsubsection{Extended Diffuse Source}
\begin{figure*}
    \centering
    \includegraphics[width=\textwidth]{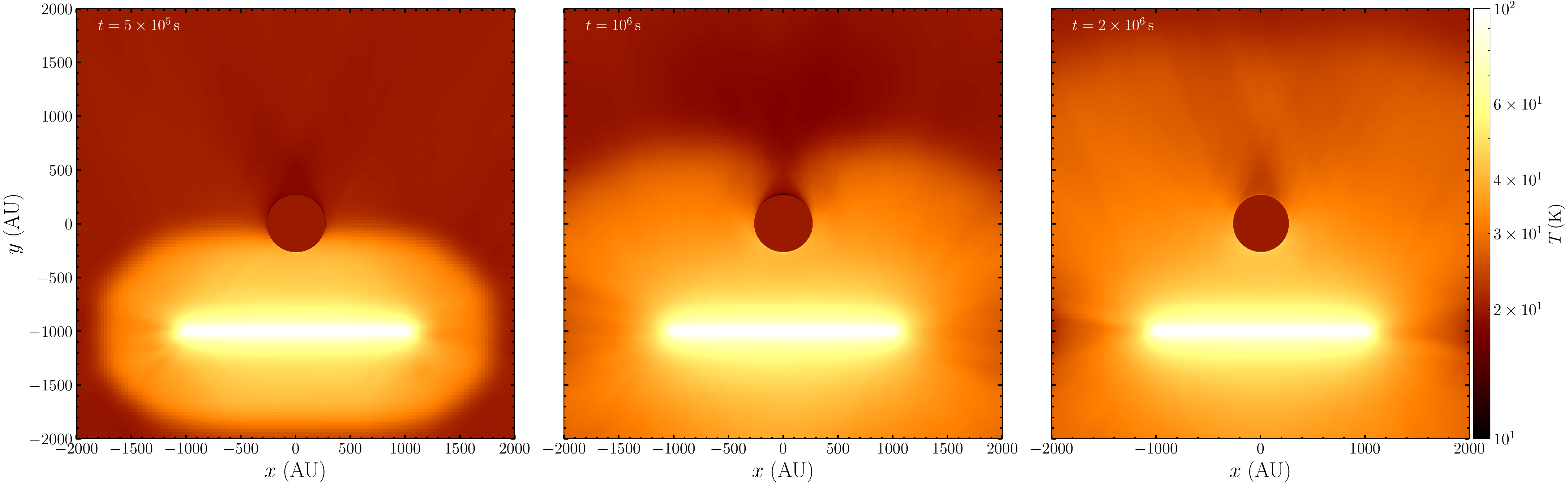}
    \caption{Time evolution of the gas temperature $T$ of an optically thick clump placed in a transparent medium, irradiated by an extended source of radiation (Equation~\ref{eq:diffuseextended}). Times correspond to 25\%, 50\% and 100\% of the light crossing time of the computational volume. Our scheme is able to capture the shadow cast by the clump very well, including subtle features such as the umbra, penumbra and antumbra.} 
    \label{fig:diffuseextended}
\end{figure*}

In our final version of the shadow test, we introduce an extended source of diffuse radiation, geometrically represented by a cylinder enclosed by half-spheres at both ends of its axis. We represent this source by the function 
\begin{equation}
    \label{eq:diffuseextended}
    j_* = 
    \begin{cases}
    j_0 \exp \left[ -\left(\frac{y_c^2+z_c^2}{2 R_*^2} \right) \right] \; & \abs{x_c} \leq l_{\mathrm{cyl}}, \sqrt{y_c^2+z_c^2} \leq 4R_* \\
    j_0 \exp \left[ -\left(\frac{x_c^2+y_c^2+z_c^2}{2 R_*^2} \right) \right] \; & \abs{x_c} > l_{\mathrm{cyl}}, \sqrt{x_c^2+y_c^2+z_c^2} \leq 4R_* \\
    0 & \text{All other } (x,y,z)
    \end{cases},
\end{equation}
where $(x_c,y_c,z_c)$ are the coordinates with respect to the center of the source at  $(0,-1000,0) \, \mathrm{AU}$, $R_* = 27 \, \mathrm{AU}$ is the characteristic size of the source, and $j_0 = L_*/(2 \pi R_*^2)^{3/2}$ where we pick a value of $L_* = 10 \, L_{\sun}$. The initial conditions for the opaque clump and ambient medium, and the fixed opacity value, are identical to those in the previous test. However, compared to this test we double the size of the computational volume to $4000 \, \mathrm{AU}$ in the $x$ and $y$ directions, in order to follow the shadow for longer times; we leave the domain size in the $z$ direction unchanged, at $2000 \, \mathrm{AU}$. We use AMR for this test, discretising the grid with a base resolution of $128 \times 128 \times 64$, and refining up to a maximum resolution of $1024 \times 1024 \times 512$. We use a refinement condition wherein blocks are tagged for refinement if a cell in the block has a relative change $\Delta f \geq 0.8$, where $\Delta f = \max \left(\Delta f_x,\Delta f_y,\Delta f_z \right)$, and -
\begin{equation}
    \Delta f_i = \frac{\left| f(i+1)-f(i-1) \right|}{\left| f(i+1)+f(i-1) \right|}, 
\end{equation}
where $f$ is the variable used for refinement, for which we use $\rho$ and $j_*$, and $i$ is the discretised cell index in the $i$'th direction where $i \in (x,y,z)$. We use 192 rays in the ray-tracer, though we obtain qualitatively identical results with 48 rays. We show the gas temperature evolution for this test in Figure~\ref{fig:diffuseextended}, at times corresponding to 25\%, 50\% and 100\% of the light crossing time of the computational volume. We can see that the optically thick clump casts a shadow when irradiated by the extended source, and our scheme is able to capture this challenging configuration of multiple converging rays quite well. We also note that subtle shadow features such as the umbra, penumbra and antumbra are noticeable, and is a testament to the ability of our scheme to handle nontrivial geometrical distributions of radiation sources. 

\label{sec:diffuseextended}

\section{Discussion}
\label{sec:Discussion}

\subsection{Comparison to FLD and $M_1$ schemes}
\label{sec:compareschemes}
The primary advantage that the scheme described in this paper offers over many other moment method based schemes is that we use a non-local closure based on the Variable Eddington Tensor (VET) obtained with a global ray-trace solution of the time-independent radiative transfer equation. The VET closure, in principle, can handle any geometrical arrangements of sources whereas local closure methods fail to propagate radiation correctly in certain situations. It is therefore interesting to compare the performance of our VET scheme to two local closures commonly-used in astrophysical codes: the flux-limited diffusion (FLD; also known as the Eddington approximation) and Moment-1 ($M_1$) closures. 

\begin{figure*}
    \centering
    \includegraphics[width=0.98\textwidth]{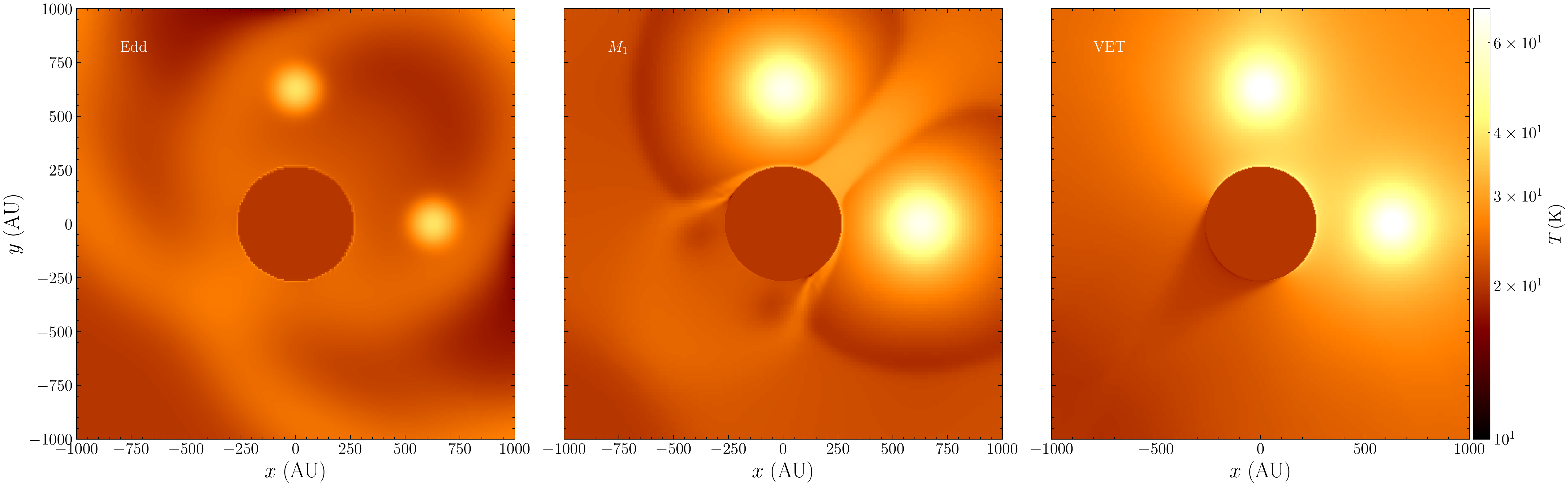}
    \caption{The numerical solution to the shadow test obtained by adopting different closures. All panels show gas temperature as a function of position at a time corresponding to one light crossing time of the computational box. The leftmost panel shows the solution obtained using the FLD (or Eddington) closure, the middle panel shows the $M_1$ closure, and the right panel shows our VET result.}
    \label{fig:Shadow_comparemethods}
\end{figure*}

To demonstrate the advantage of the VET over these local closures in complex radiation field geometries, we repeat the shadow test with spherical diffuse sources described in Section~\ref{sec:diffuseshadow} with the FLD and $M_1$ closures, and compare the results to those obtained with the VET. This setup contains multiple sources interacting in an optically thin medium, and subsequently casting a shadow, and hence represents a geometrical setup where local closures are expected to fail. To mimic the Eddington approximation, we set $\mathbfss{T} = (1/3) \mathbfss{I}$, where $\mathbfss{I}$ is the identity tensor. This is technically not identical to the FLD method, since we are still solving the equation for $\mathbfit{F}_r$, rather than determining it from the instantaneous distribution of $E_r$; however the results are expected to be qualitatively identical. To mimic the $M_1$ closure, we set 
\begin{equation}
    \mathbfss{T}=\frac{1-\chi}{2} \mathbfss{I}+\frac{3 \chi-1}{2} \hat{\mathbfit{n}} \hat{\mathbfit{n}}, 
\end{equation}
where $\hat{\mathbfit{n}} = \mathbfit{F}_r/\|\mathbfit{F}_r\|$ is the unit vector in the direction of the radiation flux and $\chi$ is the Eddington factor given by
\begin{equation}
    \chi(f)=\frac{3+4 f^{2}}{5+2 \sqrt{4-3 f^{2}}},
\end{equation}
where $f = \mathbfit{F}_r/(cE_r)$. We compute the components of $\mathbfss{T}$ with the relations above using the value of $E_r$ and $\mathbfit{F}_r$ at the beginning of the timestep (i.e. time-lagged). We show the comparison of the temperature structure after one light crossing time obtained with the three closures in Figure~\ref{fig:Shadow_comparemethods}. We can see that the Eddington and $M_1$ closure versions do not cast qualitatively correct shadows, whereas the VET version does. This demonstrates that the VET is the only closure relation that ensures the consistent propagation of radiation in non-trivial geometrical distributions of diffuse radiation sources in the presence of optically thin media. 

This leads to the important question of whether such differences could be dynamically relevant in a scientific application. This would clearly depend on the problem simulated and the potential presence of a mixture of transparent and opaque media, keeping in mind that a \textit{formal} comparison of different closure methods for a realistic numerical setup has not been performed to our knowledge. That said, the well-studied problem of a wind by trapped infrared radiation, in a gaseous atmosphere confined by a constant gravitational acceleration, could serve as a qualitative tool of comparison. 

First investigated by \citet{Krumholz_2012} with the FLD closure, this setup has been reattempted with the $M_1$ \citep{Rosdahl_2015} and VET \citep{Davis_2014} closures, and also with methods that do not rely on angular moments of the transfer equation, such as Monte Carlo Radiation Transport \citep[MCRT,][]{Tsang_2015,Smith_2020} and implicit solutions of the time-dependent radiative transfer equation \citep{Jiang_2021}. While there are broad similarities in the gas evolution with various closures, in that all schemes find that the system rearranges itself to reduce momentum transfer between radiation and gas such that the effective Eddington ratio drops from its initial value of $\sim 50$ to very close to unity, the FLD and $M_1$ closure simulations find that the steady-state Eddington ratio is slightly below unity, such that in steady state the gas remains gravitationally confined without driving a wind. On the other hand, the VET and non-moment based methods \footnote{We note that \citet{Smith_2020} compare their MCRT result with an $M_1$ closure relation version that shares the same hydrodynamics solver, and find similar discrepancies, suggesting that the adopted model for radiation transport is probably what drives the differences between these studies.} (which are expected to be at least as accurate as the VET) find that the asymptotic Eddington ratio is slightly larger than unity, leading to a slowly-accelerated wind. This suggests that, in certain problems, adopting a non-local closure and/or accurate model of radiation propagation leads to qualitatively different outcomes for the dynamical evolution of gas. There is significant scope for further work to identify, compare and quantify differences in astrophysically relevant simulation setups with the adopted treatment of radiation transport.

\subsection{Performance}
The computational cost of  \texttt{VETTAM} is very problem dependent, because the convergence of the sparse matrix solvers and the fixed-point iterations depend on the physical state of the system ($\kappa$ and $T$ distributions), which determines the stiffness of the matrix, and on the distribution of adaptive grids to MPI ranks, which determines the amount of non-local communication required. In addition, the overall performance of our scheme is highly dependent on the performance of the i) hybrid characteristics ray-tracer, for which performance and scaling capabilities are provided in \citet{Buntemeyer_2016}, and ii) solvers in the external \texttt{PETSc} library, whose performance is not directly under the control of \texttt{VETTAM}, apart from the choice of solver, preconditioner, and solver tolerance set by the user, which in certain problems can be very important. Due to these external dependencies, we do not elaborate on formal performance or scaling tests for our scheme, but rather, briefly discuss certain points relevant to performance in our scheme. 

\begin{figure}
    \centering
    \includegraphics[width=\columnwidth]{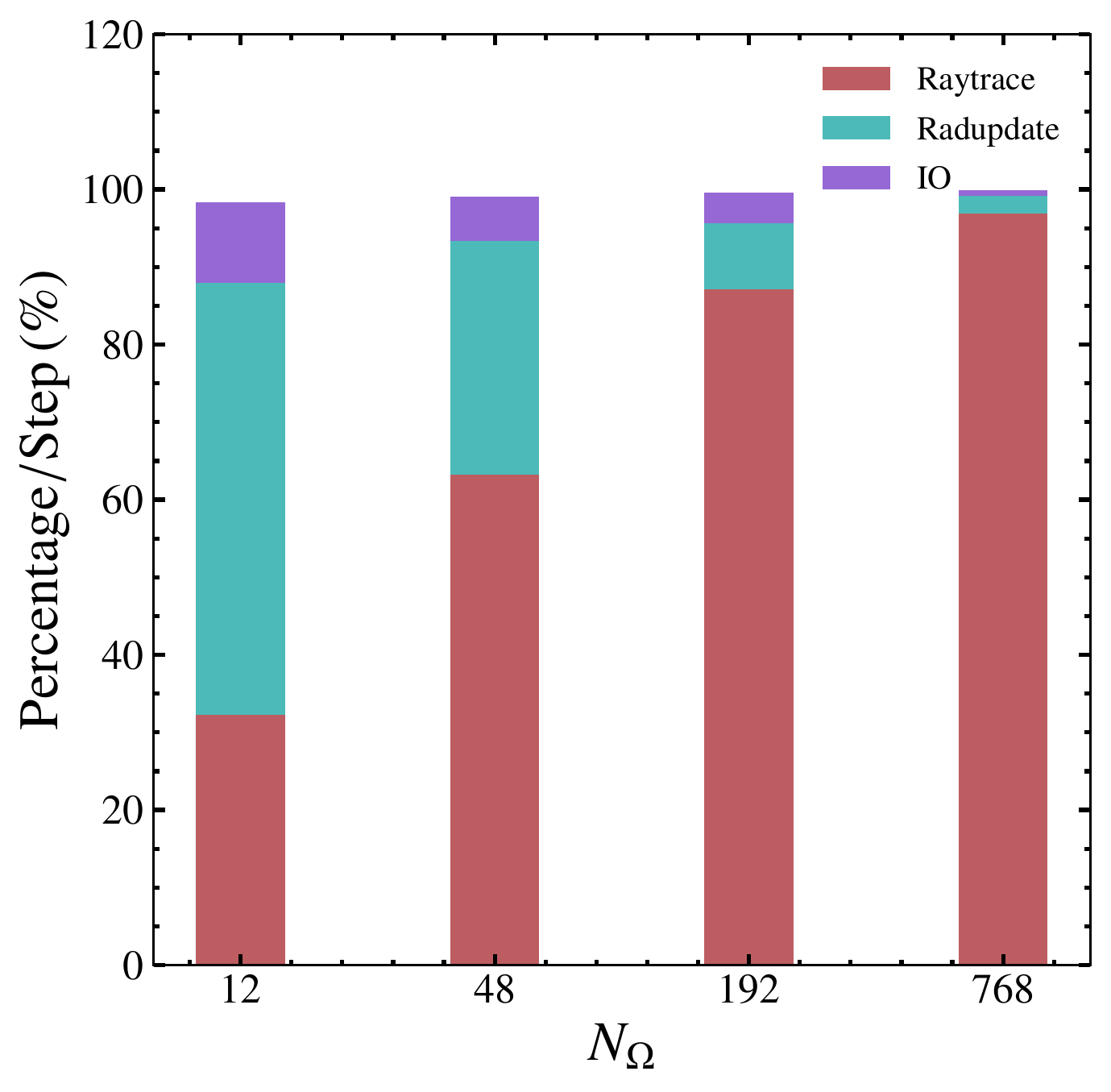}
    \caption{Breakdown of wallclock times used in the three most time-consuming units for the shadow test of Section~\ref{sec:diffuseshadow}, as a function of the number of angles $N_\Omega$ used in the ray-trace. We can see that the ray tracer occupies the largest fraction of the cost, and increasingly so at higher angular resolution.}
    \label{fig:timing_comparison}
\end{figure}

In an RHD simulation we have found that \texttt{VETTAM} occupies the largest share of computational cost. For instance, in the full RHD simulation setup of Section~\ref{sec:expandingshell}, \texttt{VETTAM} was $\sim 10$ times more expensive than the hydrodynamics update per evolution step. However, this is again, very problem dependent and this value can be higher or lower for a different problem. In general we have found the ray tracer to be the primary bottleneck to performance, though the extent to which it dominates the cost depends on the choice of angular resolution. We quantify this by running the shadow test described in Section~\ref{sec:diffuseshadow} on 144 cores, which corresponds to 3 compute nodes\footnote{All tests relevant to this section were performed on the Gadi supercomputer at the National Computational Infrastructure (NCI), Australia. Each node contains 2 x 24-core Intel Xeon Platinum 8274 (Cascade Lake) processors with 3.2 GHz CPUs per node and 192 GB of RAM, interconnected with HDR Infiniband technology in a Dragonfly+ topology.}. In Figure~\ref{fig:timing_comparison} we show the breakdown of time spent by the three most time-consuming units in the simulation-- compared for different angular resolutions for the ray-tracer (see Appendix~\ref{sec:angularres} for a comparison of results obtained for them). Since hydrodynamics was switched off for this test, it does not enter the cost breakdown. We see that even with modest angular resolutions of $48$ rays, the ray-tracer constitutes $\sim 60\%$ of the computing cost, whereas the implicit radiation update represents only $\sim 30\%$; the dominance of the ray-tracing step rises sharply at higher angular resolution. Thus, for science applications with \texttt{VETTAM}, we expect the ray-tracer to be the most expensive part of the simulation. For certain problems, we speculate that it might be possible to consider updating the Eddington tensor every few simulation timesteps, rather than at the beginning of \textit{every} timestep, without it affecting the solution significantly. This is a potential approach to alleviate the overall computational cost of the scheme in an application. 

\begin{figure}
    \centering
    \includegraphics[width=\columnwidth]{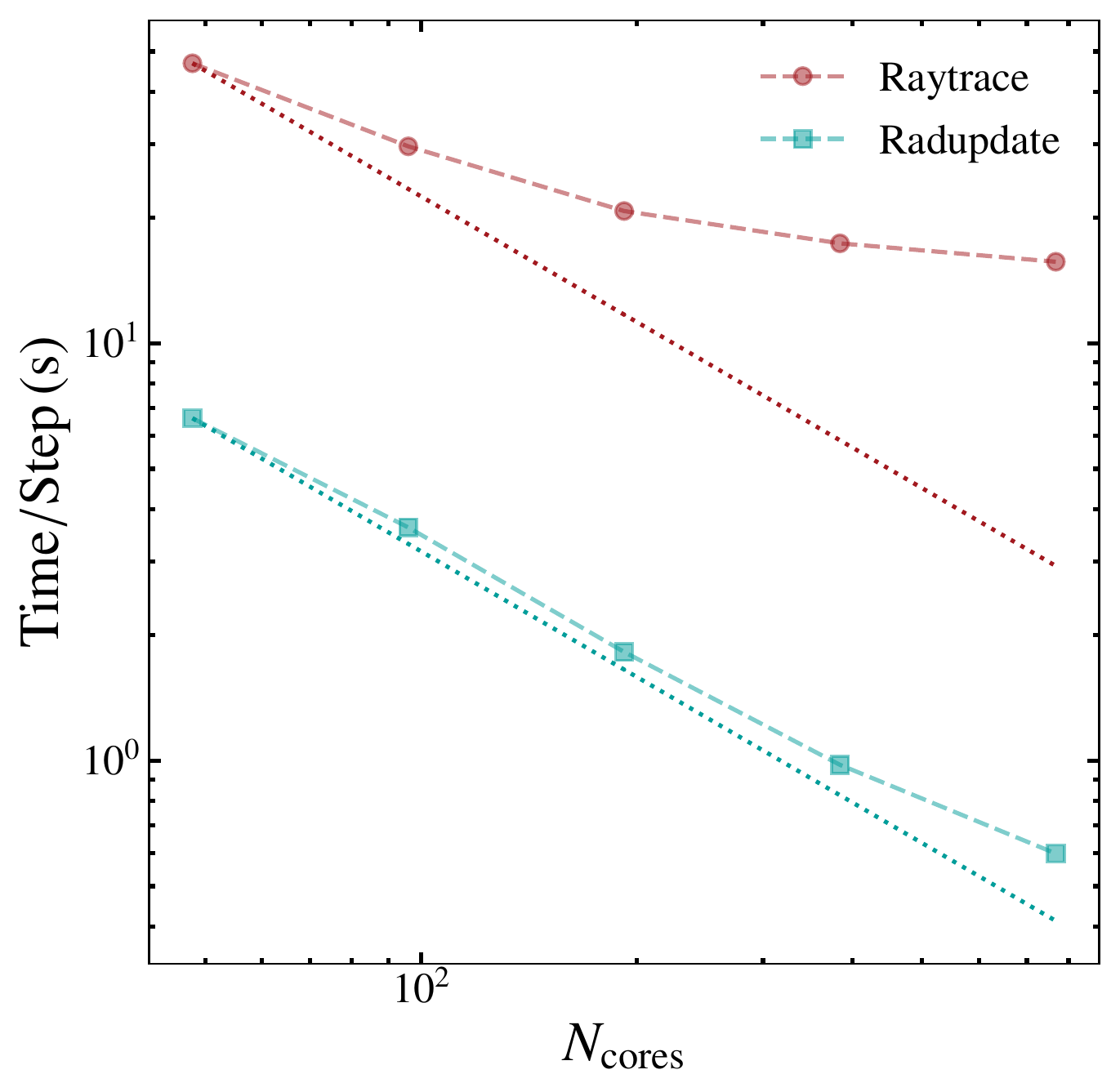}
    \caption{Strong scaling performance for the ray tracing and implicit radiation updates performed by \texttt{VETTAM} in each simulation timestep for the shadow test of Section~\ref{sec:diffuseshadow}, with $N_{\Omega} = 192$. We find reasonable strong scaling for the radiation update, but relatively poor scaling for the raytrace step due to associated communication overheads \citep[see][]{Buntemeyer_2016}.}
    \label{fig:scaling_test}
\end{figure}

Another difficulty associated with the ray-tracer has to do with its limited parallel efficiency due to inherent communication needs. This limits the strong scaling efficiency of the scheme, as the communication overheads quickly result in lower parallel efficiency, if the problem size remains fixed \citep[see,][for a discussion on this]{Buntemeyer_2016}. To quantify the scaling behaviour, we repeat the shadow test of Section~\ref{sec:diffuseshadow} with varying number of processors, keeping the AMR block structure -- and hence the total computational load -- fixed. We found in our tests that the implicit radiation moment equations update performs reasonably in strong scaling out to 768 cores, especially considering that the sparse matrix solvers have communication overheads as well. The times per evolution step in the strong scaling test for the raytrace and the radiation update are shown in Figure~\ref{fig:scaling_test}. This indicates that there is a careful choice to be made by the user to ensure that the number of blocks occupied by each processor in a parallel simulation is high enough for communication overheads not to dominate the total cost, and at the same time low enough that it satisfies the memory requirements of the ray-tracer. We aim to test, monitor and improve the performance characteristics of the scheme in the future. 

\subsection{Caveats}
In this section, we briefly mention some caveats associated with our implementation, and provide motivation for future work when applicable. 
\begin{enumerate}
    \item Our scheme is limited to the gray approximation, and this limits the usage of the scheme to problems where the spectral dependence of the radiation field is not expected to be crucial. It is, however, possible to extend our scheme with a multigroup moment method \citep{Vaytet_2011}, and this a direction for future work. 
    \item The VET is computed only once at the beginning of the timestep with the ray-trace solution of the time-independent transfer equation. However, this would not be appropriate for systems where the radiation field changes substantially on timescales much smaller than a hydrodynamic time step, so that the Eddington tensor computed at the beginning of a time step is a poor guess for its value at the end of the step. In principle one could handle such systems by iterating the Eddington tensor to convergence along with the other radiation quantities. However, this is likely computationally intractable, and it is unclear whether it would be accurate in any event -- an accurate solution would likely require choosing a time step small enough to capture the time evolution of the radiation field. 
    \item The Picard iteration scheme achieves only linear convergence, and thus the number of iterations required can be very dependent on the initial starting guess. An extension would be to implement a method to accelerate convergence, such as Anderson acceleration, or to solve the nonlinear system with a Jacobian-free Newton Krylov method; both achieve up to quadratic rates of convergence. That being said, Picard iteration generally has a greater radius of convergence compared to Newton methods. At present there is little reason to optimise the iterative solve, since the total cost is dominated by the ray-trace, but as we improve the ray-trace and relieve this constraint, we also intend to improve the iterative scheme. 
    \item \texttt{VETTAM} solves the equations of RHD in the mixed frame formulation by expanding the lab frame opacities in terms of the comoving frame opacities to $\mathcal{O}(v/c)$ \citep{Mihalas_1982}. This formulation is poorly-suited to line radiation transport, and hence we are limited to continuum radiation. 
    \item Our scheme is first-order in space and time, and while this maintains stability and simplicity, it can be diffusive in certain problems due to the associated truncation errors (for instance in Section~\ref{sec:Radiating_Pulse} for the streaming regime). A direction for future work would be the development of higher order implicit Godunov methods to treat the RHD equations that resolves the associated difficulties of maintaining monotonicity in an implicit higher-order method \citep[for a discussion, see Section 4.2,][]{Sekora_2010}. 
    
\end{enumerate}

\section{Summary}
\label{sec:Summary}
In this paper we describe Variable Eddington Tensor-closed Transport on Adaptive Meshes (\texttt{VETTAM}), a multidimensional RHD algorithm that solves the mixed-frame radiation moment equations closed with a non-local Variable Eddington Tensor (VET) computed through a formal solution of the time-independent radiative transfer equation. \texttt{VETTAM} is, to our knowledge, the first ever implementation of a VET closure scheme that can handle Adaptive Mesh Refinement (AMR). We have coupled our implementation to our own private version of the \texttt{FLASH} hydrodynamics code \citep{Fryxell_2000} which uses the \texttt{PARAMESH} library for AMR \citep{Macneice_2000}. Our scheme has been designed to handle continuum radiation transport mediated by dust, although other radiative mechanisms such as photoionisation \citep[e.g.][]{Aubert_2008,Kuiper_2020}, or alternative transport phenomena such as cosmic-ray or neutrino transport, can be performed with slight modifications \citep[e.g.,][]{Jiang_2018}. We use a finite-volume Godunov method with an HLLE Riemann solver for the radiation moment equations with an implicit backwards-Euler time update that allows us to evolve the system at the hydrodynamic timescale. In addition, we treat the coupled nonlinear radiation-matter energy exchange term through a fixed-point Picard iteration method, that effectively linearises the backwards-Euler update, for which we use the sparse matrix solving capability provided by the \texttt{PETSc} library. The formal solution to the time-independent transfer equation for computing the VET is performed through the hybrid characteristics ray-tracing scheme implemented by \citet{Buntemeyer_2016}. We carry out a comprehensive suite of tests to demonstrate that our scheme works correctly in different regimes of radiation transport, and can handle the coupling between radiation and hydrodynamics correctly. We also demonstrate through a test that other commonly used local closure methods such as FLD or $M_1$ yield unphysical radiation fields in certain physical scenarios, the dynamical effects of which are difficult to predict and poorly explored \citep[e.g.,][]{Davis_2014}. However, implicit VET methods are computationally more expensive than other closures due to the inherent communication needs of the ray-tracing scheme and matrix inversion algorithms. We argue, however, that the computational cost benefits offered by AMR in our scheme will be crucial in applications that require both spatial accuracy and computational efficiency. Currently we are using \texttt{VETTAM} to study the effects of reprocessed infrared radiation pressure on super-star cluster formation in dense molecular clouds \citep[e.g.,][]{Skinner_2015,Tsang_2018}, and intend to explore other relevant applications in the near future. 

\section*{Acknowledgements}
S.~H.~M would like to thank Yan-Fei Jiang and Shane W.~Davis for discussions that assisted in the progress of the project. S.~H.~M would also like to thank M.~Aaron Skinner and Eve Ostriker for discussions on their implementation of the M1 method in Athena, and Anna L.~Rosen for information on the shadow test. 
C.~F.~acknowledges funding provided by the Australian Research Council through Future Fellowship FT180100495, and the Australia-Germany Joint Research Cooperation Scheme (UA-DAAD). M.~R.~K.~acknowledges funding from the Australian Research Council through its \textit{Discovery Projects} and \textit{Future Fellowship} funding schemes, awards DP190101258 and FT180100375. RK acknowledges financial support via the Emmy Noether and Heisenberg Research Grants funded by the German Research Foundation (DFG) under grant no.~KU 2849/3 and 2849/9. We further acknowledge high-performance computing resources provided by the Leibniz Rechenzentrum and the Gauss Centre for Supercomputing (grants~pr32lo, pn73fi, and GCS Large-scale project~22542), and the Australian National Computational Infrastructure (grants~ek9 and~jh2) in the framework of the National Computational Merit Allocation Scheme and the ANU Merit Allocation Scheme.

\textit{Software}: \texttt{PETSc} \citep{PetscConf,PetscRef}, \texttt{NumPy} \citep{numpy}, \texttt{SciPy} \citep{scipy}, \texttt{Matplotlib} \citep{matplotlib}, \texttt{yt} \citep{yt}. This research has made use of NASA's Astrophysics Data System (ADS) Bibliographic Services.

\section*{Data Availability}
No new data was produced from this study. \texttt{VETTAM} has been implemented in our own private forked version of \texttt{FLASH}, and would be shared on reasonable request to the corresponding author.



\bibliographystyle{mnras}
\bibliography{VETTAM} 




\appendix
\section{Hyperbolic Wavespeeds for Radiation Subsystem}
\label{sec:wspeed}

\subsection{HLLE Wavespeed Correction}
\label{sec:wavespeed}
In this section we demonstrate that our approach to estimating wavespeeds for describing the HLLE Riemann fluxes at cell interfaces produces the correct solution in the diffusion limit, and show why the wavespeed correction as described in Section~\ref{sec:hllefluxes} is necessary, especially with Adaptive Mesh Refinement (AMR). The fundamental issue is that using a maximum/minimum wavespeed $(\lmax,\lmin)$ for the Riemann flux based on the characteristic speed obtained from the eigenvalues of the streaming limit radiation moment equation leads to a scheme that is too diffusive when the optical depth across a cell $\tau \ga 1$. This is because the numerical diffusive flux due to the HLLE Riemann solver can be much larger than the physical radiative flux \citep{Audit_2002}. To circumvent this issue, one can modify the wavespeed in a manner described in Section~\ref{sec:hllefluxes}. We demonstrate below that this improves the solution considerably, even when the optical depth per cell is close to unity. 

\begin{figure*}
    \centering
    \includegraphics[width=0.98\textwidth]{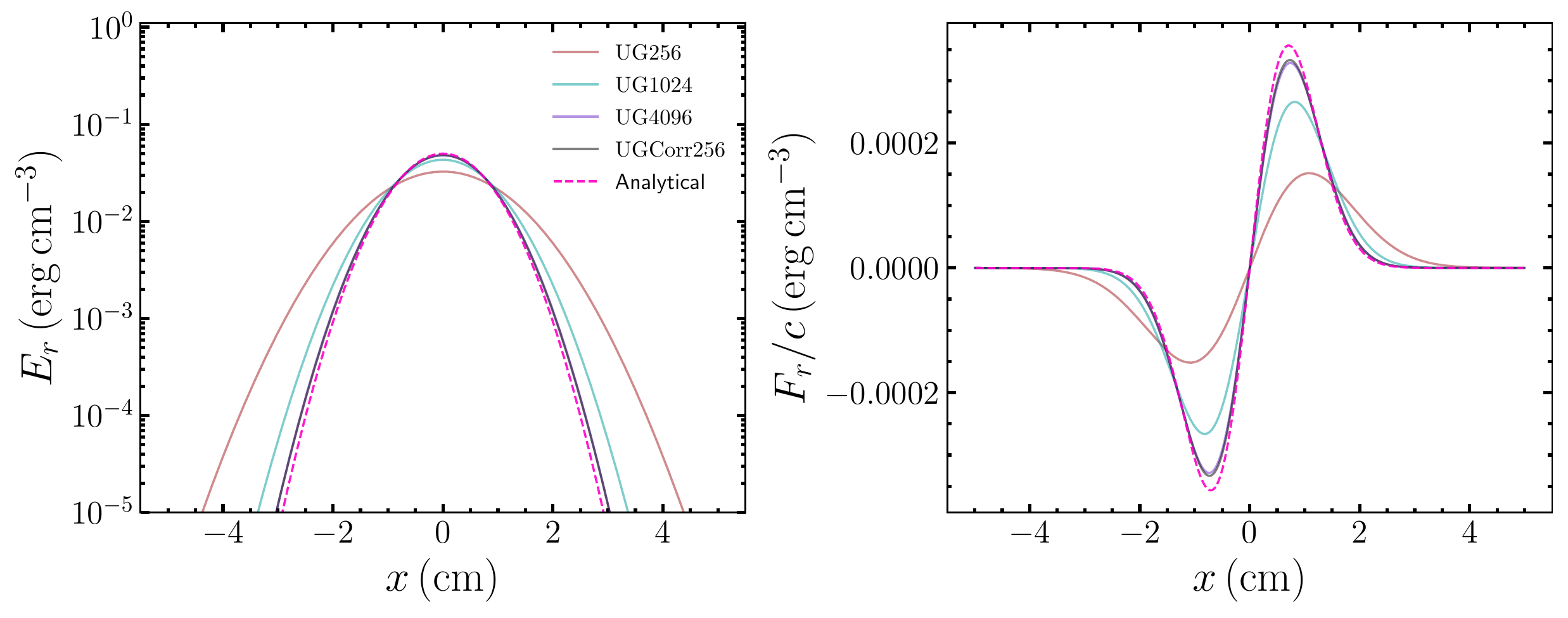}
    \caption{The numerical solution for the radiation energy ($E_r$) and flux ($F_r$) of the diffusing radiating pulse test of Section~\ref{sec:Radiating_Pulse} obtained for a uniform grid (UG) with resolutions $N_x$ (optical depth/cell $\tau$) of 256 ($\tau \sim 1.5$), 1024 ($\tau \sim 0.4$) and 4096 cells ($\tau \sim 0.1$) without the HLLE wavespeed correction described in Section~\ref{sec:hllefluxes} (solid coloured lines), and that for a resolution of 256 cells with the wavespeed correction (solid black line). The dashed pink line shows the exact solution provided in Section~\ref{sec:Radiating_Pulse}. We see that the numerical solution is too diffusive without the wavespeed correction, and requires high resolution to converge to the right solution, whereas the version with wavespeed correction remains close to the right solution even at lower resolutions.}
    \label{fig:J14Problem}
\end{figure*}

To demonstrate why wavespeed correction is necessary, we set up a problem identical to the weak equilibrium pulse test described in Section~\ref{sec:Radiating_Pulse}, without performing the wavespeed correction, for three different uniform grids of 256, 1024 and 4096 cells. For comparison, we also perform a version of the test at a resolution of 256 cells with the wavespeed correction enabled. Using the parameters of the test, the optical depths per cell with this setup are $\tau \approx 1.5, 0.4$, and $0.1$ respectively for the three resolutions, making the cells only marginally optically thick. However, even in this case, we show in Figure~\ref{fig:J14Problem} that the solution without the wavespeed correction is significantly more diffusive, and only converges to the analytical solution when the photon mean free path is well resolved. On the other hand, we find that we can obtain similar accuracy at a resolution of $256$ cells with the wavespeed correction. We note that the setup is only marginally optically thick, and the effect of the correction would be even larger at higher cell optical depths. The test above verifies that the wavespeed correction is a robust approach to obtain the right solution in the diffusion limit. 

\subsection{Wavespeed Correction at AMR Level Boundaries}
\label{sec:wavespeedamr}
\begin{figure}
    \centering
    \includegraphics[width=\columnwidth]{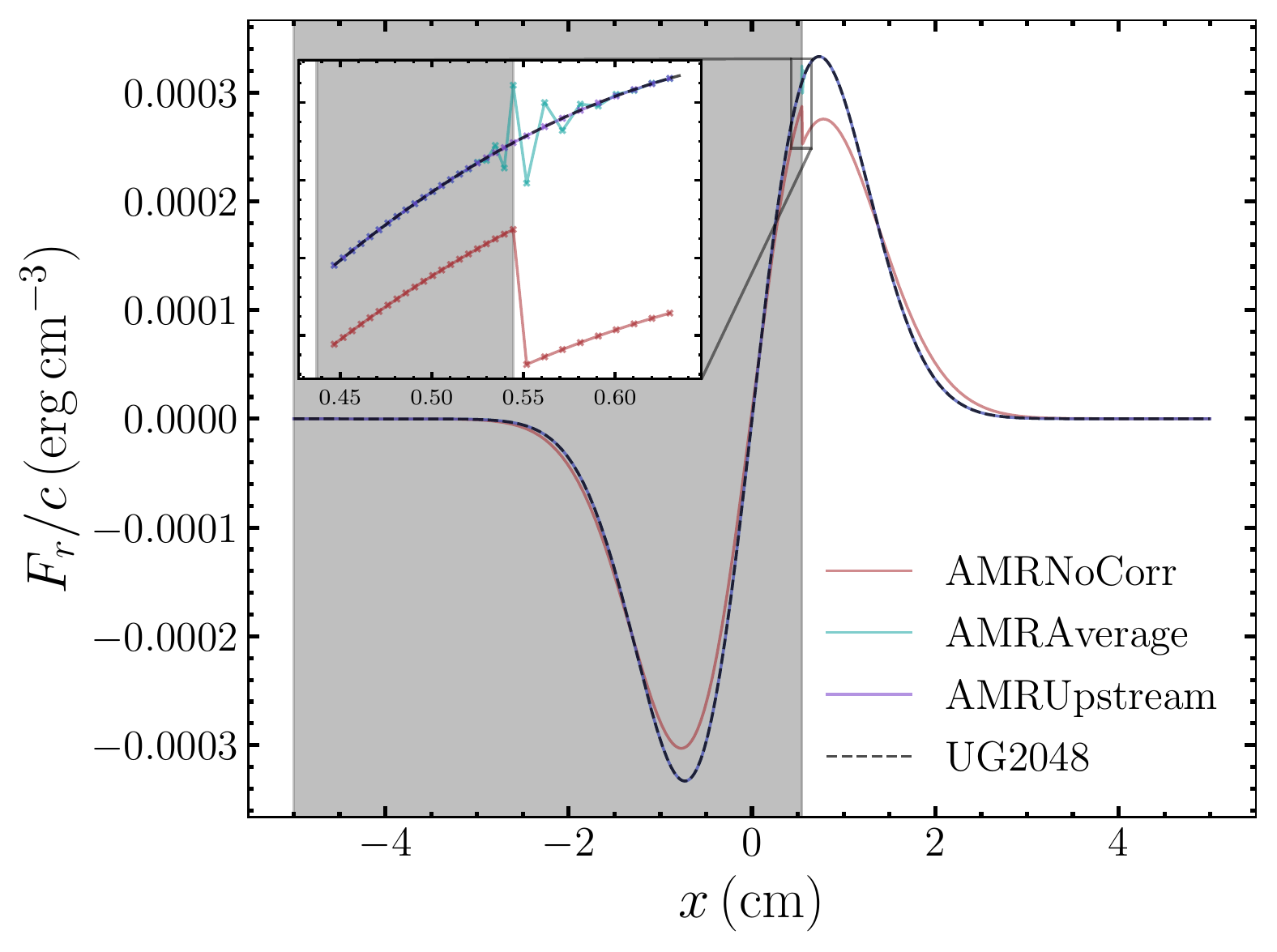}
    \caption{The numerical solution for the radiation flux ($F_r$) for the weak equilibrium diffusion test of Section~\ref{sec:Radiating_Pulse} in a domain with a fine (shaded) to coarse (unshaded) AMR level boundary, with the fine cells at a resolution of $\Delta x \sim 0.0048 \, \mathrm{cm}$ and the coarse cells at $\Delta x \sim 0.0097 \, \mathrm{cm}$. We show four variations of the test - i) without the HLLE wavespeed correction described in Section~\ref{sec:hllefluxes} (AMRNoCorr) ii) corrected wavespeeds using an arithmetic average optical depth at the interface for the correction factor (AMRAverage), iii) corrected wavespeeds using the optical depth of the cell upstream to the wave propagation direction (AMRUpstream), and iv) a uniform grid version with wavespeed correction at the resolution of the finer level of the AMR domain (UG2048). The inset zooms in on the cells near the level interface to demonstrate the presence of discontinuities (oscillations) in the AMRNoCorr (AMRAverage) version. We can see that AMRUpstream version provides a smooth and accurate solution throughout the domain.}
    \label{fig:AMRCorrection}
\end{figure}

We next demonstrate the generalisation of our wavespeed correction to interfaces where the left and right cells have different widths (i.e. at level interfaces). As described in Section~\ref{sec:hllefluxes}, for interfaces between cells at the same AMR level, we use the arithmetic average of the optical depth of the cells sharing the interface to compute the correction factor. On the other hand, for an interface at an AMR level boundary, where one neighbouring cell is finer than the other, we use the upstream value of the optical depth for the correction factor. To show why this change is necessary, we repeat the weak equilibrium diffusion pulse test of Section~\ref{sec:Radiating_Pulse} with a base grid resolution of 1024 cells. We enforce a simple fixed refinement condition, refining the region $x<0.5$ by a factor 2 in cell width, leaving the region $x>0.5$ at the base resolution, which leads to a single level boundary in the domain. We perform four variations of the test for comparison - i) without the HLLE wavespeed correction, ii) using wavespeeds corrected with a correction factor computed from the arithmetic average optical depth at the interface, iii) using wavespeeds corrected with a correction factor computed from the optical depth of the cell upstream to the wave propagation direction, and iv) a uniform grid (UG) version of the problem where the entire domain is resolved by 2048 cells, corresponding to the resolution of the finer level of the AMR domain. We show the numerical solution we obtain for the radiation flux $F_r$ for these four cases in Figure~\ref{fig:AMRCorrection}, shading the region of the domain that is refined to a higher AMR level. We show only $F_r$ as we found $E_r$ to be relatively smooth for this test even with AMR, though in some other tests we found discontinuities at AMR levels in $E_r$ as well.

We can clearly see from Figure~\ref{fig:AMRCorrection} that there is a sharp discontinuity at the level interface for the version without the wavespeed fix (AMRNoCorr). This discontinuity can be explained by the fact that the numerical diffusivity, which dominates over the physical diffusivity without the wavespeed correction, changes discontinuously across the coarse-fine interface, and this leads to a mismatch in the solution at the interface. The wavespeed correction using the average optical depth at the interface (AMRAverage) alleviates this discontinuity, but instead produces spurious oscillations near the interface. On the other hand, using the optical depth of the upstream cell for the correction factor (AMRUpstream) leads to a smooth solution that matches that obtained by the uniform grid at the finer AMR level grid resolution (UG2048). We also experimented with several other possible methods of choosing the optical depth at the interface of AMR level transitions, and found that the upstream version led to the best results.

\section{Dependence on Angular Resolution of Ray Tracer}
\label{sec:angularres}
\begin{figure*}
    \centering
    \includegraphics[width=0.98\textwidth]{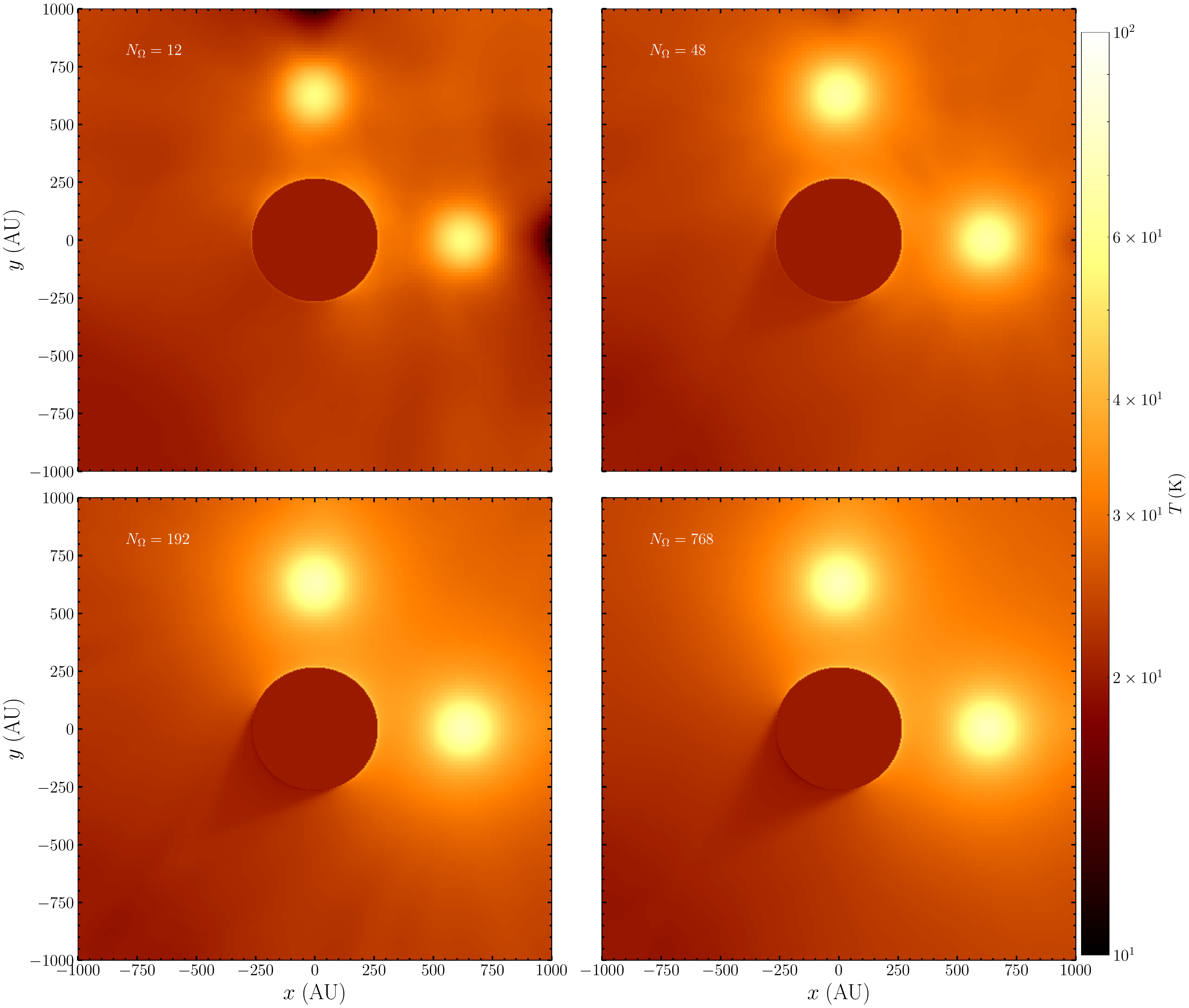}
    \caption{Comparison of the temperature structure obtained in the test described in Section~\ref{sec:diffuseshadow} for varying numbers of angles ($N_\Omega$) used in the ray-tracer to compute the VET.}
    \label{fig:Shadow_comparenSide}
\end{figure*}

We derive the variable Eddington tensor (VET) in our scheme from a ray trace-based solution to the time-independent radiative transfer equation, and the accuracy of the VET depends on the accuracy of our ray tracing method. The most important parameter that controls this accuracy is the angular resolution $N_{\Omega}$, i.e. the number of discrete angles over which the transfer equation is solved. As outlined in Section~\ref{sec:ComputeVET}, we use the \verb|HEALPix| scheme to discretise the unit sphere into equal-area pixels, with a base resolution of 12 angular pixels, and further levels differing by a factor of 4 in the number of angles (i.e. $N_{\Omega} = 48, 192, 768, \ldots$). To evaluate the effects of varying $N_{\Omega}$, we repeat the shadow test described in Section~\ref{sec:diffuseshadow} with four different angular resolutions: 12, 48, 192, and 768 angles. We show the resulting temperature structures after a light crossing time of the computational box in Figure~\ref{fig:Shadow_comparenSide}. We find that all four angular resolutions produce reasonable temperature structures, but, as expected, the shadow becomes increasingly sharp with larger $N_\Omega$; however, even for $N_\Omega = 48$ a clear umbra is visible. While the dependence on angular resolution will in general be problem-dependent, this comparison provides confidence that reasonable results can be obtained even with relatively modest angular resolutions.


\bsp	
\label{lastpage}
\end{document}